\documentclass[12pt]{article}
\usepackage{graphpap,epsfig,amssymb,graphicx,textcomp}
\usepackage[utf8]{inputenc}
\usepackage[english]{babel}
\begin {document}

\begin {center}
{\bf {\Large Nonequilibrium phase transition in spin-S Ising ferromagnet
 driven by Propagating and Standing magnetic field wave.}}
\end {center}
\begin {center}{\bf{ $\bf Ajay ~Halder^\dagger$ and $\bf Muktish ~Acharyya^\star$}
\\{Department of Physics,}\\{Presidency University}\\
{86/1 College Street, Kolkata-700073, India}
\\{$\bf^\dagger ajay.rs@presiuniv.ac.in$}
\\{$\bf^\star muktish.physics@presiuniv.ac.in$}}

\end {center}
\vskip 0.2cm
\begin {abstract}

The dynamical response of spin-S (S=1, 3/2, 2, 3) Ising ferromagnet to the plane propagating wave 
, standing magnetic field wave and uniformly oscillating field with constant frequency 
are studied separately in two dimensions by extensive Monte Carlo simulation. 
Depending upon the strength of the magnetic field and the value of the spin state
 of the Ising spin lattice two different dynamical phases are observed. For 
a fixed value of $S$ and the amplitude of the propagating magnetic field wave the 
system undergoes a dynamical phase transition from propagating phase to pinned phase
 as the temperature of the system is cooled down. Similarly in case with standing 
magnetic wave the system undergoes dynamical phase transition from high temperature phase 
 where spins oscillates coherently in alternate bands of half wavelength of the 
standing magnetic wave to the low temperature pinned or spin frozen phase. 
For a fixed value of the amplitude of magnetic field oscillation the transition 
temperature is observed to decrease to a limiting value as the value of spin $S$ is increased.
 The time averaged magnetisation over a full cycle of the magnetic field oscillation 
plays the role of the dynamic order parameter. A comprehensive phase boundary 
is drawn in the plane of magnetic field amplitude and dynamic transition temperature. 
It is found that the phase boundary shrinks inwards for high value of spin state $S$.
 Also in the low temperature (and high field) region the phase boundaries are closely spaced. 

\end {abstract}



\newpage

{\noindent \bf 1. Introduction}
\vskip 0.5cm

Ising model has long been used to understand the behaviour of ferromagnetic system 
in thermodynamic equilibrium as well as in nonequilibrium conditions.
 Specially dynamical response\cite{rmp,ijmpc} of a ferromagnet under nonequilibrium situations 
is quite interesting. Mainly the nonequilibrium dynamic phase transition and the hysteretic response 
characterise any ferromagnetic system driven by time dependent magnetic field. 
Dynamic specific heat\cite{divcp}, relaxation time\cite{relax}
 and the relevant length scale near the transition point\cite{divln} appear to 
diverge near the transition point. 
These along with tricritical behaviour\cite{trict1,trict2}, hysteresis loss\cite{hyst}
 etc. show similarity of such dynamical behaviour
 with the well known equilibrium thermodynamic phase transition. Ferromagnetic 
behaviour under the influence of oscillating magnetic field was also studied
 in continuous systems and in anisotropic systems as well;
 such as: off-axial dynamic phase transition in Heisenberg model\cite{anisoHei} and in XY model\cite{xy}, 
multiple (surface and bulk) dynamic transition in classical Heisenberg model\cite{mulHei}, 
 dynamic phase transition in kinetic spin 3/2 Blume-Capel model\cite{bc}
 and in Blume-Emery-Griffiths model\cite{beg} and so on.
 Behaviour of mixed spin systems\cite{deviren1,temizer,vatans,ertas,shi} have also been studied recently.
 Mainly Mean field approximation and the Monte-Carlo simulation techniques
 were used in these studies. 
 Blume-Capel model has widely been used in various anisotropic
 ferromagnetic systems\cite{stauffer,jain,deserno,lara,deviren2} 
such as $S=1$, $S={3 \over 2}$, $S={5 \over 2}$ etc. to study the bicritical/tricritical 
behaviours in phase transitions. Using renormalization group theory in Midgal-
Kadanoff approximation\cite{yunus} three dimensional $S={3 \over 2}$ Ising system has been
 studied recently and a very rich phase diagram is obtained. General spin BC
 model was studied using meanfield approximation\cite{general}. Competing 
behaviours of the metastable states in BC model\cite{fiig,manzo} were also studied using dynamic 
Monte Carlo and numerical transfer matrix method.

 More interesting dynamical phase transitions between various dynamical phases  
have been observed\cite{JMMM1,JMMM2,appb,ajay,bcwav} recently in Ising ferromagnets as well as in BC ferromagnets driven by
 propagating and standing magnetic waves. 
In these situations magnetic field varies both in space and time throughout the lattice. 
Effect of waves on transition temperature and phase boundaries were mainly studied here. 

 Instead of specific spin system, this would be interesting to know the general 
behaviour of a $S$-spin ferromagnet under intense magnetic field wave. How the 
transtion temperature and the dynamical phase boundary in a ferromagnet
 change with the number of spin states under various types of externally
 applied magnetic waves may reveal interesting facts about ferromagnetic systems.

In this paper, we have investigated the response of an $S$-spin Ising ferromagnet 
 in the presence of propagating and standing magnetic field wave using 
Monte-Carlo simulation. The paper is organised as follows: 
The model and the MC simulation technique are discussed in Sec. II, the numerical 
results are reported in Sec. III and the paper ends with a summary in Sec. IV.

\vskip 0.6cm

\noindent {\bf 2. Model and Simulation}
\vskip 0.5cm

The {\it time dependent} Hamiltonian of a two dimensional Ising ferromagnet, 
having $S$ numbers of spin states is represented by, 
\begin{equation}
 H(t) = -J\Sigma\Sigma' s^z(x,y,t) s^z(x',y',t) - \Sigma h^z(x,y,t)s^z(x,y,t) .
\end{equation}
Here $s^z(x,y,t)$ is the $z$ component of the Ising {\it S-state unit spin variable} at lattice site $(x,y)$ at time $t$. 
The summation  $\Sigma'$ extends over the nearest neighbour sites $(x',y')$ of a given site $(x,y)$.
 $J(>0)$ is the {\it ferromagnetic spin-spin interaction strength} between the nearest neighbour pairs of spin. 
The value of $J$ is considered to be uniform over the whole lattice, for simplicity.
 The externally applied {\it magnetic field}, $h^z(x,y,t)$, at site $(x,y)$ at time $t$,
 has the following forms for Propagating wave, Standing wave and uniformly oscillating field respectively, 
\begin{equation}
 h^z(x,y,t)= h_0 cos \{2\pi (ft-\frac{x}{\lambda})\} 
\end{equation}
\begin{equation}
 h^z(x,y,t)= h_0 sin (2\pi ft) sin (2\pi \frac{x}{\lambda})
\end{equation}
\begin{equation}
 h^z(x,y,t)= h_0 cos (2\pi ft)
\end{equation}
Here $h_0$ and $f$ represent respectively \textit{the field amplitude and the frequency}
 for the propagating magnetic wave, standing magnetic wave as well as uniformly oscillating field,
 whereas $\lambda$ represents \textit{the wavelength} for both the waves.
%These waves are {\it not} the usual spin waves in ferromagnets, rather they are externally applied into the ferromagnetic system.
The propagating wave propagates along the $X$-direction and the modulation of the 
standing wave is also taken along the same direction.

A model of an $L\times L$ square lattice of Ising spins having 
{{\textsl{periodic boundary conditions}}}, applied at both directions, is considered here.
Such boundary conditions preserve the translational invariances in the system.
 The spins have unit magnitude and S states. For eg. a 3-state spin has values of spin variable
 $s^z=+1,\ 0\ and\ -1$ whereas a 4-state spin has values 
$+\frac{3}{2}$, $+\frac{1}{2}$, $-\frac{1}{2}$ and $-\frac{3}{2}$
for $s^z$ and so on. The results are simulated using {\it Monte Carlo Metropolis single spin flip algorithm} with 
{\it parallel} updating rule\cite{springer}. 
The initial phase is chosen as the high temperature random disordered phase,
 where all the $s^z$ values of the ferromagnetic spins have equal probabilities.
 The {\it Metropolis rate} of spin flip at temperature $T$ is given by,
\begin{equation} 
W((s^z)_i\to (s^z)_f) = Min [exp(\frac{-\Delta E}{k_BT}),1] 
\end{equation}
 where $\Delta E$ is the energy change due to spin flip from $i$-th state to $f$-th and 
$k_B$ is the Boltzman constant. Updating of $L^2$ spin states in an 
$L\times L$ square lattice constitute the unit time step called 
{\it Monte Carlo Step per Spin} (MCSS). The units of the applied magnetic field and 
the temperature are $J$ and $J/k_B$, respectively. 

In the present study we have taken $L=100$.
The system is cooled down slowly in small steps ($\Delta T=0.02$) from the high-temperature phase,
i.e., the dynamical disordered phase, before reaching any dynamical 
steady phase at lower temperature $T$. This particular choice of system size
is a compromise between the computational time and finite size effect. The detail study of finite size analysis (for different system sizes) is going
on which requires a huge computational time and will be reported after having
the results.

\newpage

\noindent {\bf 3. Results}

\vskip 0.5 cm
3.1. \underline {Propagating field wave:}
\vskip 0.2cm

To study the nonequilibrium behaviour of S-spin Ising ferromagnet in 2D we have
 taken a lattice of dimension ($100\times 100$). Propagating magnetic 
field waves having different values of field amplitude $h_0$ but fixed wavelength $(\lambda=25\ lattice\ units(lu))$ 
and frequency $(f=0.01\ MCSS^{-1})$ are allowed to pass through the system. The dynamical 
quantities at any temperature $T$ is calculated when
 the system has achieved steady state. For this we have kept 
the system at constant temperature $T$ for a sufficient long 
time ($100000\ MCSS$) i.e. through $1000$ complete
 cycles of magnetic oscillations while discarding initial 
(or transient) $500$ cycles and taking average over 
the remaining $50000\ MCSS$. We have identified a phase transition 
between high temperature {\it symmetric phase}
 and low temperature {\it symmetry-broken phase}.
 The {\it order parameter} for the phase transition is defined as the average 
magnetisation of each spin over a complete cycle of magnetic field oscillation,
 i.e. $Q=f\times \oint M(t)dt$, where $M(t)$ is the instantaneous magnetisation per spin state at time $t$.
 At very high temperature the order parameter has a low value which means that 
the spins are symmetrically distributed over all of their S states. At high temperature 
 thermal energy of spins exceeds their mutual interaction energy and hence in varying magnetic 
field these spins flips more easily along the direction of field oscillation depending upon the strength of the magnetic field.
 As a result this phase propagates coherently alongwith the propagating wave. 
This propagation of spins have no relation with the conventional spin waves in ferromagnets. 
This phase may also be called as {\it propagating phase}.
 When the system is cooled down and the magnetic energy alongwith the thermal 
energy becomes insufficient to overcome the mutual interaction strength between any pair of spins 
 this symmetric distribution of spin states breaks and the absolute value of magnetisation begins to grow  
as the temperature falls below the transition temperature. Thus the value of the order parameter
 becomes high. The variations of the {\it instantaneous magnetisation per spin state}, measured as
$M(t)={{1} \over {L^2}} \sum_i S_i^z(x,y,t)$, with time in symmetric phase and 
symmetry-broken phase are shown in figs.1 for $3$-state and $7$-state spins respectively.
 In symmetric phase magnetisation varies around {\it zero} value resulting in very 
low average whereas it varies around a {\it nonzero} value in symmetry-broken phase.
 Figs.2 show the variation of different 
dynamical quantities at steady state; such as: order parameter $(Q)$, time derivative of 
the order parameter $\frac{dQ}{dT}$, variance of the order parameter $(V=L^2(\langle Q^2\rangle-\langle Q\rangle ^2))$, and the 
dynamic heat capacity $(C_v=\frac{dE}{dT})$ for $3$-state, $5$-state and $7$-state spins respectively.
 Order parameter takes on nonzero value as the system cools and becomes unity below certain
 transition temperature defining the dynamic 
phase transition. The transition is detected by the sharp variations of 
$\frac{dQ}{dT}$, $V$ and $C_v$ at the transtion temperature $(T_d)$.
 It is observed that the transition temperature {\it decreases} and approaches a minimum value
as the number of spin states {\it increases}.
 Fig.3 shows the {\it dynamic phase diagrams $(T_d\ vs\ h_0)$} for 3-state, 5-state and 7-state spins
 respectively which shrinks inwards for greater number of spin states $S$.
These also manifest that the transition occurs at lower temperature ($T_d$) for higher values
 of field amplitude ($h_0$) as we usually observe in other studies with 2-state Ising spins. 
Here, it may be noted that the typical size of the errorbar of
the data in Fig-2 is around 0.03. From the peak (or dip) positions of various
quantities the transition temperature was determined. Here the maximum possible
error in estimating the transition temperature will be 0.02 (this is the value
of $\Delta T$ by which the temperature of the system is reduced). So, the 
data shown in the phase diagram (in Fig-3) involves the error of size 0.02
in the estimation of transition temperature. 

\vskip 0.2 cm
3.2. \underline {Standing field wave}:
\vskip 0.2cm

The dynamic phase transition is also observed in Ising ferromagnet having S number of spin states
 driven by standing magnetic wave. Here also the lattice size is $100\times 100$.
 In our study we have chosen fixed value for wavelength of the wave which is $\lambda=24.5\ lu$ so that there
 are exactly 8 loops of magnetic oscillation in a  lattice. The frequency of magnetic oscillation
 is taken as $(f=0.01\ MCSS^{-1})$. Steady state dynamical behaviour is observed 
by keeping the system in constant temperature for a sufficently long time $200000\ MCSS$
 i.e. for through $2000$ complete oscillations of magnetic field. Various dynamical 
quantities are calculated after discarding initial $1000$ cycles of oscillations 
and then taking average over the remaining $100000\ MCSS$. At steady state 
we have observed a dynamical phase transition between high temperature symmetric phase and low 
temperature symmetry-broken phase similar to that with propagating wave.
 The {\it order parameter} for the transition is also defined in a way similar to that of the 
propagating wave. At sufficiently high temperature
 all the S states of spin have equal probability and hence the
 average magnetisation per spin $M(t)$ at any time $t$ is nearly 
zero, which is shown in fig.4b \& fig.4d for 3-state and 7-state spins respectively,
 whereas at low temperature thermal energy of spins
 are much less and the probability of spin flip is mainly magnetic field driven. If the 
magnetic field amplitude becomes low the mutual interaction energy between any pair of spins
 does not allow them to respond to the variations of magnetic field coherently and hence 
the value of magnetisation per spin at any time $t$ takes nonzero value as
 shown by the fig.4a. \& fig.4c. respectively for 3-state and 7-state spins. 
Unlike the situation in propagating magnetic field wave, in standing magnetic
 wave there is a local variation of field amplitude; zero at the nodes and 
maximum at the antinodes. So at nodes of the standing wave the dynamics of 
the spin states are always thermally driven but at antinodes the value of field amplitude
 may have the effect on spin flip at relatively lower but above the transition temperature.
% As a result there is nonuniform distribution of different spin states over the lattice near above the transition temperature.
 Above the transition temperature alternate bands of spins ($s^z=+1\ or\ -1$)
 forming standing wave is observed. This wave is not the usual spin wave in ferromagnets. 
Near the boundaries of the standing wave bands there are almost equal 
population of all S-states of spin.
 Below the transition temperature field energy as well as thermal energy 
is insufficient to alter any spin state and all the ferromagnetic spins orient themselves in a particular 
direction giving rise to symmetry-broken or spin-frozen phase with nonzero magnetisation per spin. 
 Thus the order parameter varies continuously from zero at high temperature to unity
 below the transition temperature as shown in fig.5a. for 3-state, 5-state and 7-state spins
 respectively. The transition is detected by observing the sharp variation of 
$\frac{dQ}{dT}$, $V$ and $C_v$ with temperature near the dynamic transition temperature $T_d$. 
These variations are shown in fig.5b., fig.5c. and fig.5d. respectively.
 All of these variations show that the transition temperature {\it decreases} and approaches a minimum value as the
 number of spin states {\it increases}. We have also drawn three comprehensive 
phase boundaries for 3-state, 5-state and 7-state spins in $T_d$-$h_0$ plane (fig.6.). 
It is observed here also that the phase boundary shrinks inwards for greater number
 of spin states. The phase boundaries also show that the transition temperature
 decreases as  the magnetic field amplitude increases similar to previous results 
obtained with 2-state Ising spins.
Here, it may be noted that the typical size of the errorbar of
the data in Fig-5 is around 0.03. From the peak (or dip) positions of various
quantities the transition temperature was determined. Here the maximum possible
error in estimating the transition temperature will be 0.02 (this is the value
of $\Delta T$ by which the temperature of the system is reduced). So, the 
data shown in the phase diagram (in Fig-6) involves the error of size 0.02
in the estimation of transition temperature. 

\vskip 0.2cm
3.3. \underline {Uniformly Oscillating field}:
\vskip 0.2cm

We have also checked the nonequilibrium phase transition in S-state Ising ferrromagnets 
under uniformly oscillating magnetic field, where there is no spatial variation of the 
magnetic field throughout the Ising spin lattice. We kept the size of the lattice and 
the frequency of magnetic field oscillation similar to the above mentioned studies with 
propagating and standing magnetic wave. Steady state behaviour is studied as previously
 mentioned. It is observed that the system undergoes phase transition from high temperature 
symmetric phase to low temperature symmetry-broken phase depending on the values of 
the magnetic field amplitude $h_0$ and the number of spin states $S$. At sufficiently high 
temperature spins are uniformly distributed over all of its states which is considered as 
the initial state of the spin system. As the temperature is cooled this uniform distribution
 over all the states is no longer observed, rather spins orient themselves along the 
direction of the externally applied magnetic field and average magnetisation per spin $(M(t))$
 follows the magnetic oscillation. The value of order parameter is zero as a result. 
 This is shown in fig.7b. and fig.7d. for 3-state and 7-state spins respectively.
 At temperature below the transition temperature the probability of spin flip is much 
low and the spins freeze in a fixed direction parallel to the magnetic oscillation
 and $M(t)$ takes nonzero value as shown in the fig.7a. and fig7c. respectively.
 The value of order parameter hence also becomes nonzero. 
The variations of $Q$, $\frac{dQ}{dT}$, $V$ and $C_v$ with temperature $T$ are shown in 
the fig.8a., fig.8b., fig.8c. and fig.8d. respectively. The variations observed in these figures 
show that the transition temperature {\it decreases} and approaches a minimum value as the number of spin 
states {\it increases}. We have also drawn comprehensive phase boundaries for 
3-state, 5-state and 7-state spins in $T_d$-$h_0$ plane (fig.9.) which shrink inwards for greater 
number of spin states $S$. These phase boundaries also reveals that the 
transition temperature decreases as the field amplitude increases.
Here, it may be noted that the typical size of the errorbar of
the data in Fig-8 is around 0.03. From the peak (or dip) positions of various
quantities the transition temperature was determined. Here the maximum possible
error in estimating the transition temperature will be 0.02 (this is the value
of $\Delta T$ by which the temperature of the system is reduced). So, the 
data shown in the phase diagram (in Fig-9) involves the error of size 0.02
in the estimation of transition temperature.

\vskip 0.6cm

\noindent{\bf 4. Summary:}
\vskip 0.2cm

The dynamics of $S$-state Ising ferromagnet in the presence of propagating magnetic wave,
 standing magnetic wave and uniformly oscillating magnetic field has been studied here 
using Monte Carlo simulation with parallel updating rule for the spin states. Two 
distinct dynamical phases namely: {\it symmetric} phase 
and {\it symmetry-broken} phase are observed depending on the values of temperature,
 strength of the magnetic field and the number of states of the Ising spins for all kinds of magnetic excitations. 
For a fixed field amplitude the dynamic transition temperature decreases as the number of spin states increases.
 Transitions occur also at lower temperature for higher magnetic field amplitudes for a fixed number of spin states.
 The symmetric phase propagates coherently with the propagating magnetic wave whereas in case of standing
 magnetic wave alternate bands of spins oscillate out of phase forming standing waves of spin bands in symmetric phase. 
For uniformly oscillating field the spins oscillates coherently with the magnetic field.
 Below transition temperature the spins orient in some fixed direction (either up or down)
 and are not affected by the magnetic fields, yielding the maximum for the absolute value of magnetisation. 
The transitions are detected by observing the variations of $Q$, $\frac{dQ}{dT}$, $V$ and $C_v$ 
with $T$. Transition temperatures, found from the peaks in the $V$-$T$ or $C_v$-$T$ curves are
 employed to draw the phase boundaries.

 Qualitative nature of the nonequilibrium phase transition driven by different magnetic excitations
are more or less same in case of multiple state Ising spins. The transition temperature decreases  
towards a limiting value (in the limit of very large number of states)
with greater number of spin states. 

The question naturally arises here, what is the reason of considering
different types of external driving magnetic field. Although the qualitative
nature of the phase boundary is same for all different types of magnetic fields,
the morphological structures of the dynamical spin configurations are different
for different kinds of magnetic field. These are shown explicitly in the
figures. For propagating magnetic field Fig-10 shows the coherent
propagation of spin bands. The variations of spin configuration with temperature
, field amplitude and the values of spin S , are shown in Fig-11. As a 
contrast, the coherent spin propagation is absent in the case of standing
magnetic wave. In this case, the change in values (of the spins)
 of the alternate bands
is observed here. This is shown in Fig-12. Here also, the variations of
spin configuration with temperature, field amplitude and values of spin S,
are shown in Fig-13. The uniform (over the space) and time 
dependent (sinusoidal) magnetic field does not have any spatial variation
at any instant of time. So, one should not expect any dynamical pattern
in the spin configuration. This is shown in Fig-14. The variations of 
spin configuration ({\it without any significant pattern}) with temperature
, field amplitude and values of spin S, are shown in Fig-15.

The present study, though looks pedagogical, has a motivation with experimental
background. Recently, the site diluted Blume-Capel model was 
studied\cite{expt1} by meanfield renormalization group analysis with good
agreement of the experimental phase diagram of Fe-Al alloy. In our case,
studied here, this can be generalised for S=1 Ising ferromagnet.
This coherent propagation of spin bands can be experimentally studied
by time resolved magneto optic Kerr (TRMOKE) effect. We believe, that this
has a significant role in the field of spintronics and magmonincs\cite{bader}.
The magnetic behaviours of core-shell magnetic nanoparticles has an important
role in the magnetism research as well as in the technology. The properties
of magnetism have been studied\cite{zeng} in bimagnetic ($FePt/MFe_2O_4
(M=Fe,Co))$ core-shel nanoparticles. The nonequilibrium phase transition
has been studied \cite{polat} by Monte Carlo simulation in spherical core-shell
(s=3/2 core and s=1 shell) under time dependent (uniform over space) magnetic
field. We propose to study the dynamic responses of core-shell magnetic
nanoparticles in the presence of magnetic field having spatio temporal
variation in the form of propagating and standing magnetic wave.

\newpage

\noindent{\bf V. References:}

\vskip 0.6cm
\footnotesize{
}
\newpage
%%%FIG-1%%%%%%%%%%%%%%%%%%%%%%%%%%%%%%%%%%%%%%%%%%%%%%%%%%%%%%%%%%%%%%%%

\begin{figure}[h]
\begin{center}
\begin{tabular}{c}
\resizebox{6cm}{6cm}{\includegraphics[angle=0]{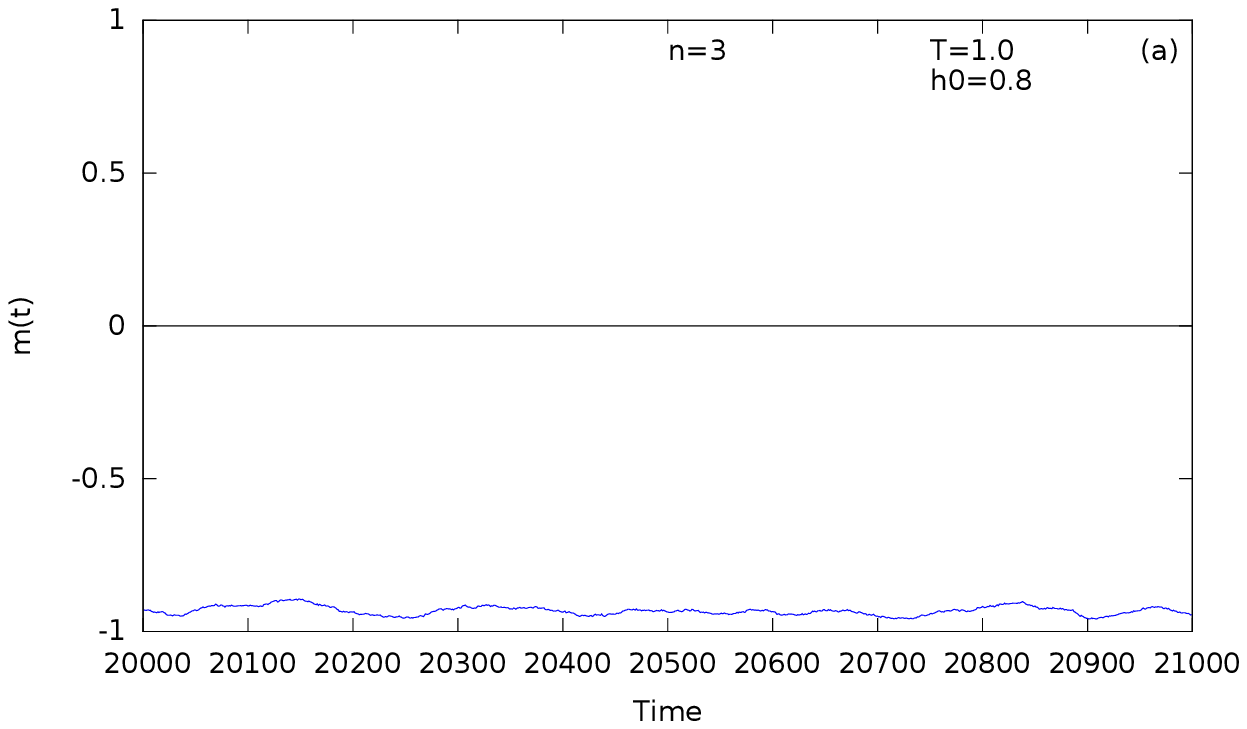}}
\resizebox{6cm}{6cm}{\includegraphics[angle=0]{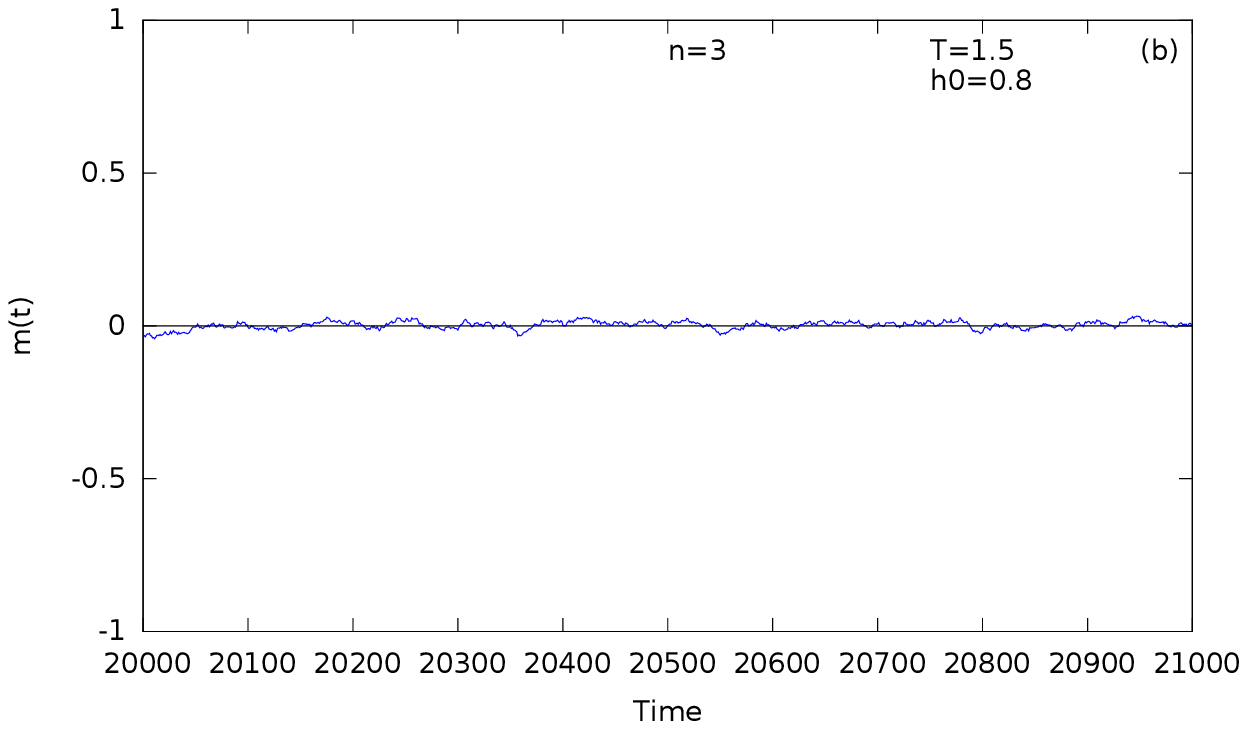}}
\\
\resizebox{6cm}{6cm}{\includegraphics[angle=0]{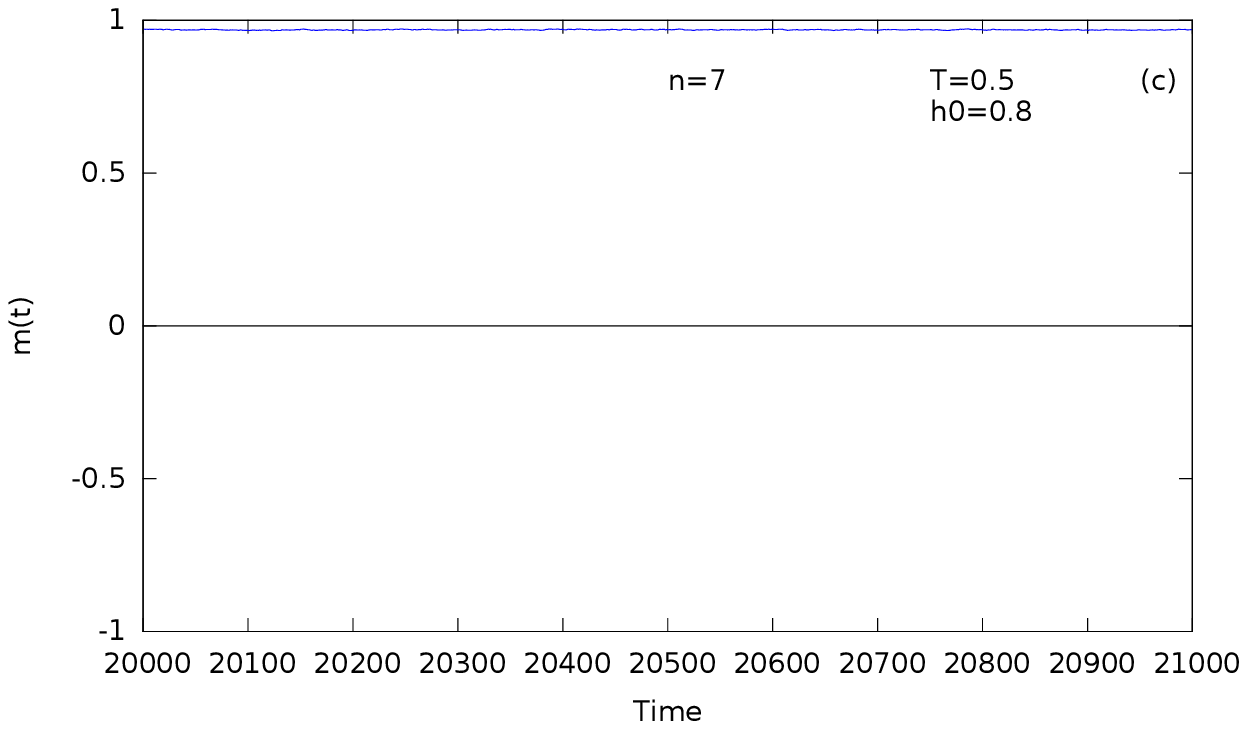}}
\resizebox{6cm}{6cm}{\includegraphics[angle=0]{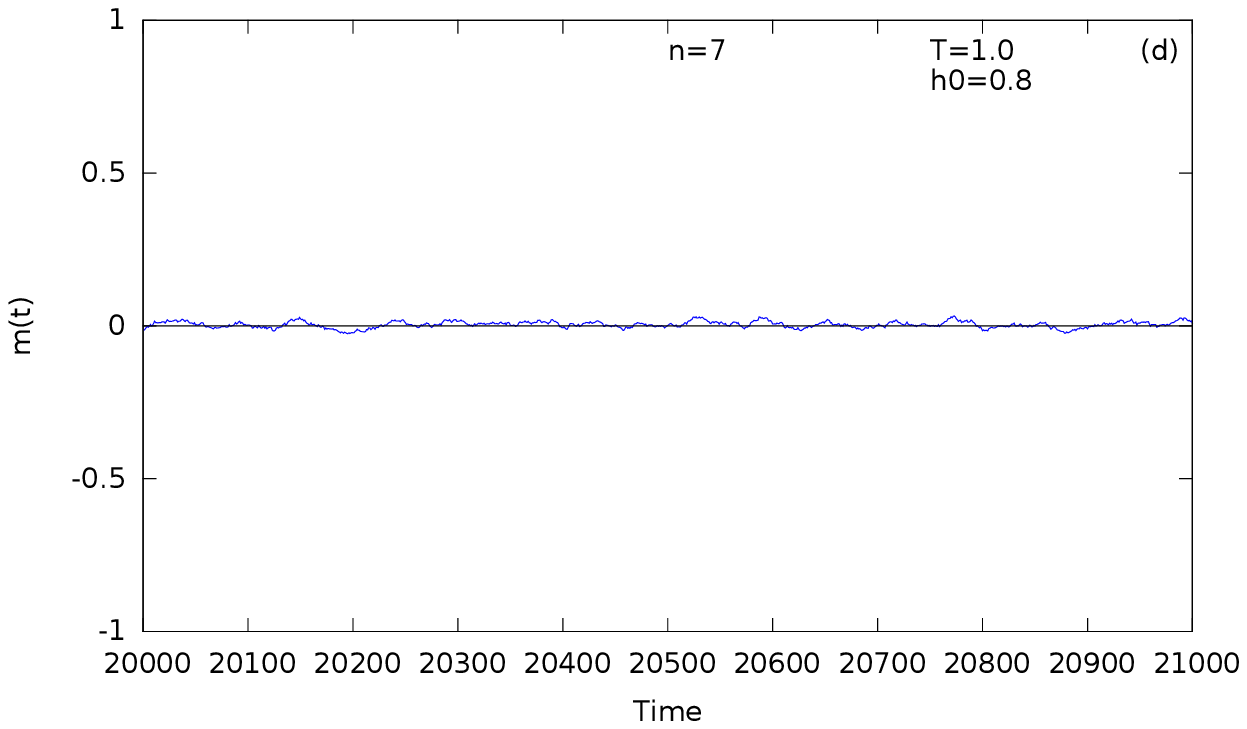}}
          \end{tabular}
\caption{\footnotesize{Variation of magnetisation with time at constant
 temperature ($T$) and field amplitude ($h_0$) for propagating wave.
 Fig.1a \& fig.1b. represent {\it 3-state} spin whereas
 fig.1c. \& fig.1d. represent {\it 7-state} spin. Frequency $f=0.01$.}}

\end{center}
\end{figure}

%%%%%%%%%%%%%%%%%%%%%%%%%%%%%%%%%%%%%%%%%%%%%%%%%%%%%%%%%%%%
\newpage
%%%%%%%%%%%%%%FIG-2%%%%%%%%%%%%%%%%%%%%%%%%%%%%%%%%%%%%%

\begin{figure}[h]
\begin{center}
\begin{tabular}{c}
\resizebox{6cm}{6cm}{\includegraphics[angle=0]{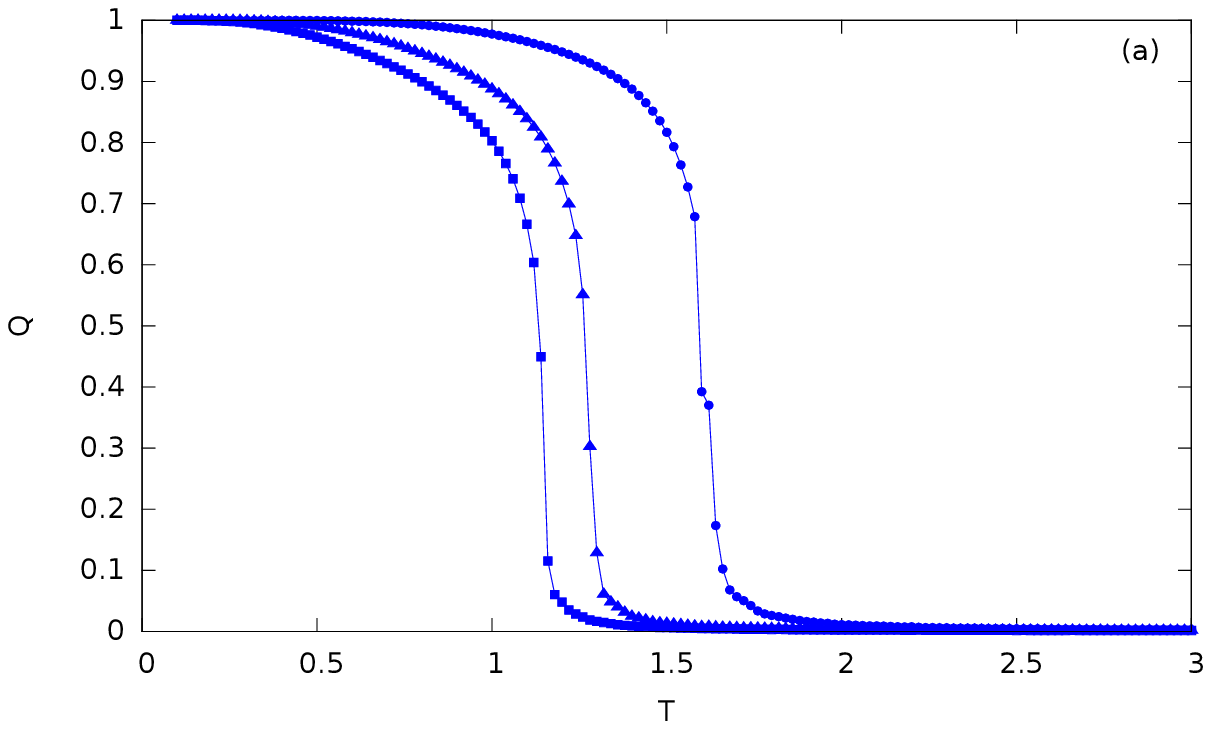}}
\resizebox{6cm}{6cm}{\includegraphics[angle=0]{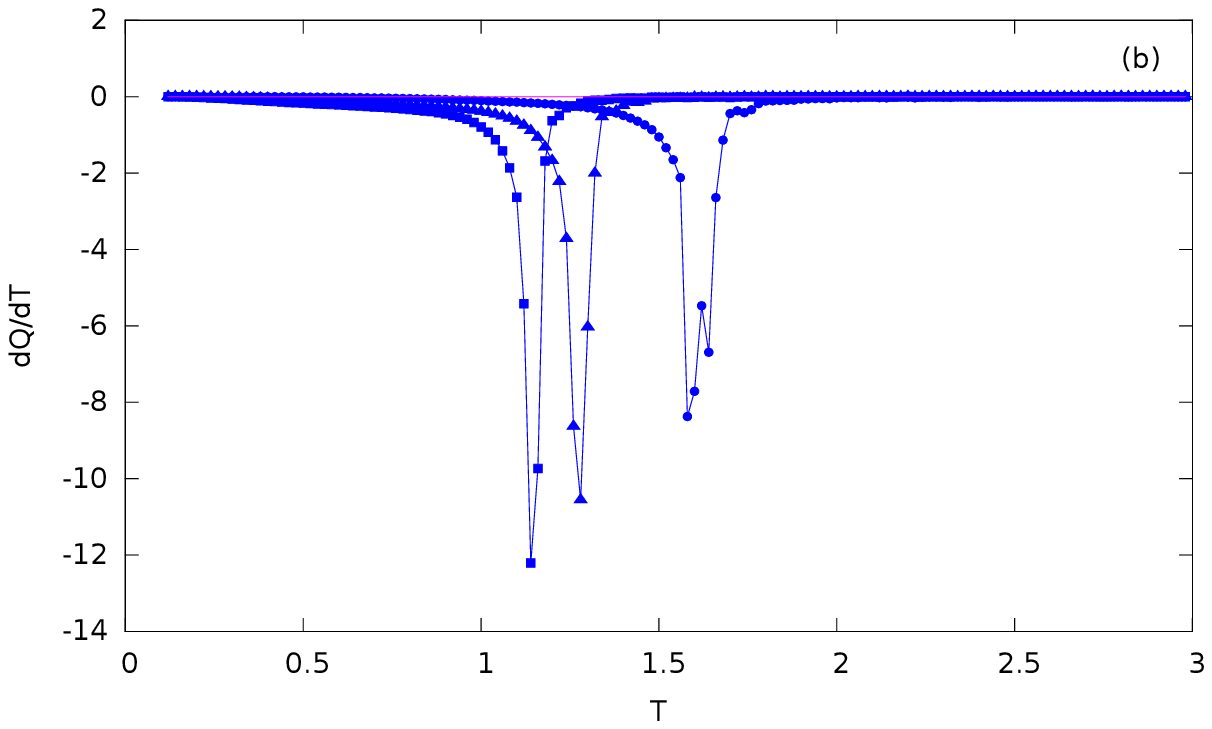}}
\\
\resizebox{6cm}{6cm}{\includegraphics[angle=0]{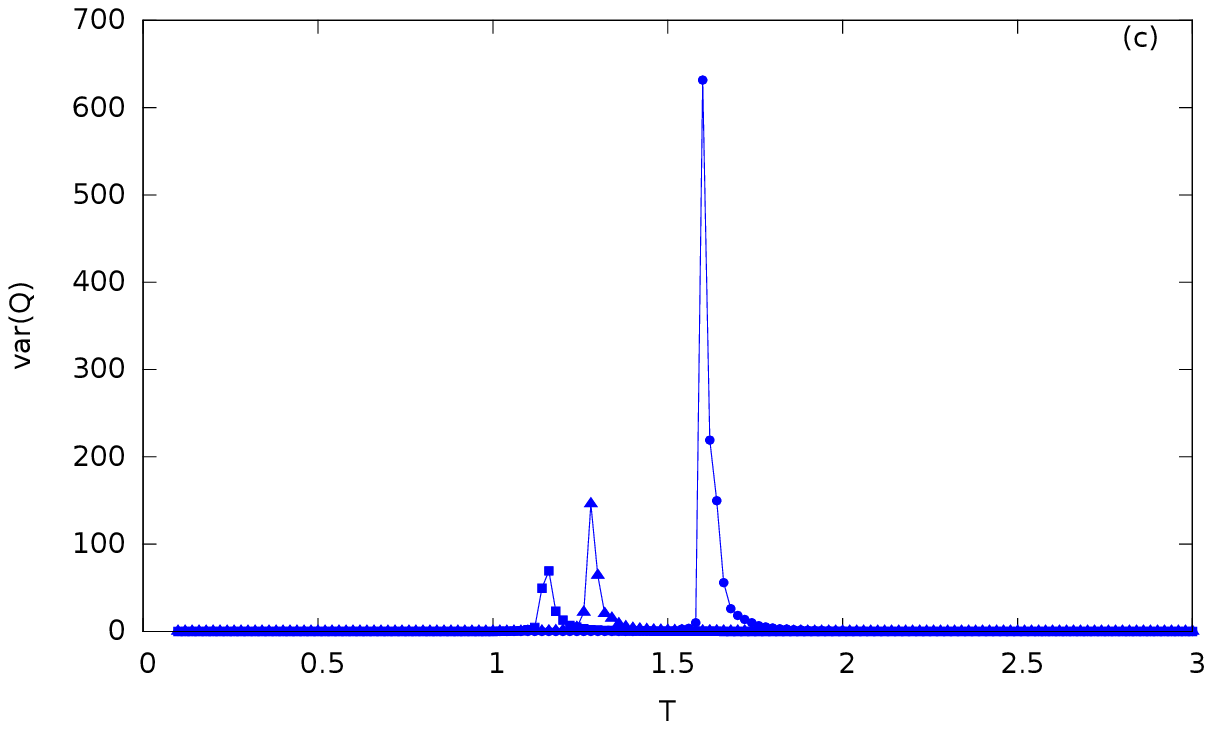}}
\resizebox{6cm}{6cm}{\includegraphics[angle=0]{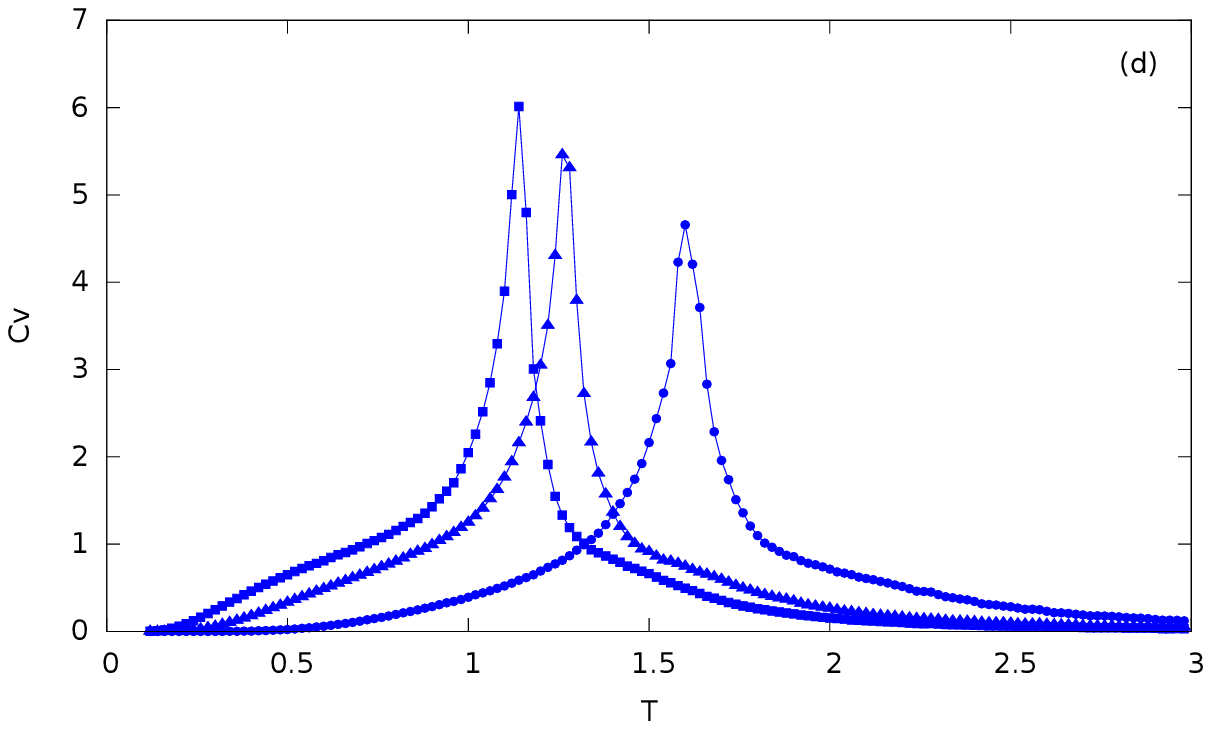}}
          \end{tabular}
\caption{\footnotesize{Temperature variation of (a) $Q$, (b) $\frac{dQ}{dT}$,
 (c) $V$, (d) $C_v$ for {\it 3-state} (circle), {\it 5-state} (uptriangle) and 
{\it 7-state} (square) spins for constant field amplitude ($h_0=0.2$)
 and frequency ($f=0.01$) of propagating wave.}}

\end{center}
\end{figure}

%%%%%%%%%%%%%%%%%%%%%%%%%%%%%%%%%%%%%%%%%%%%%%%%%%%%%%%%%%%%%%%%%%%%%%%%%%%%%%%%%%
\newpage
%%%%%%%%%%%%%%%%%%FIG-3%%%%%%%%%%%%%%%%%%%%%%%%%%%%%%%%%%%%%%%%%%%

\begin{figure}[h]
\begin{center}
\begin{tabular}{c}
\resizebox{6cm}{6cm}{\includegraphics[angle=0]{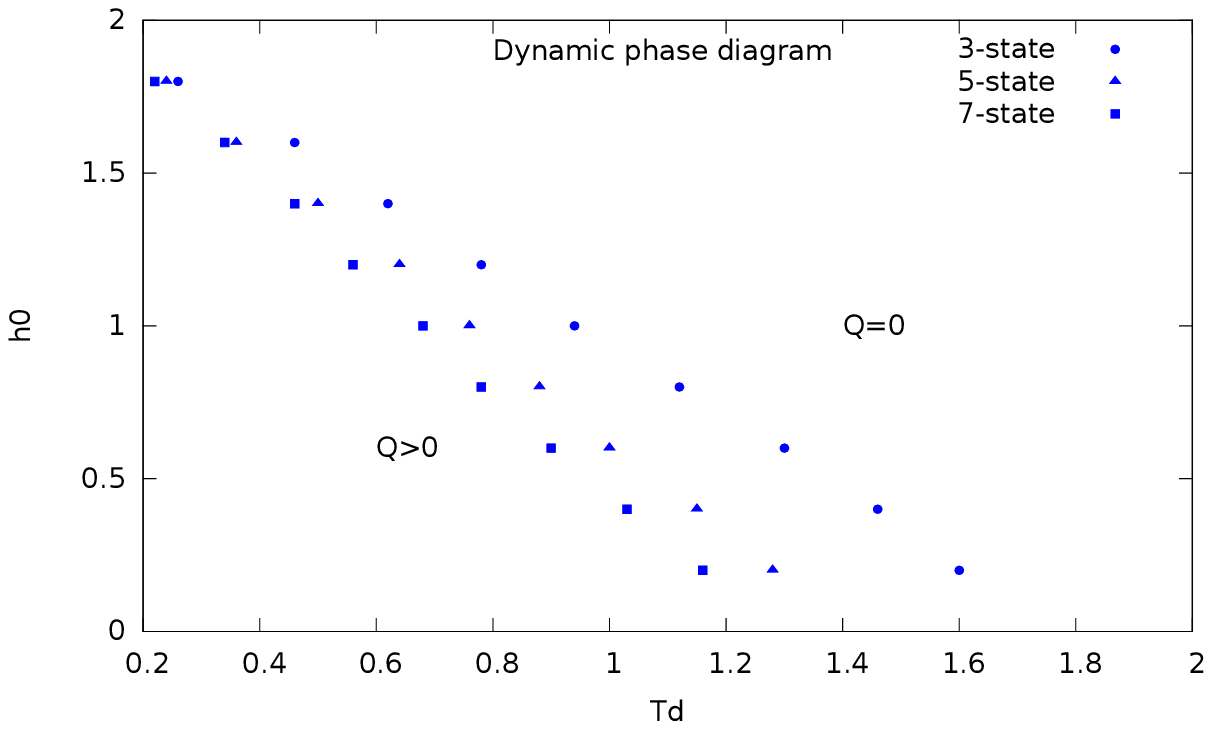}}
\\
          \end{tabular}
\caption{\footnotesize{Phase diagram in $T_d$-$h_0$ plane for {\it 3-state} (circle), {\it 5-state} (uptriangle) and 
{\it 7-state} (square) spins for propagating wave.}}

\end{center}
\end{figure}

%%%%%%%%%%%%%%%%%%%%%%%%%%%%%%%%%%%%%%%%%%%%%%%%%%%%%%%%%%%%%%%%%%%%%%%%%%%%%%
\newpage
%%%%%%%%%%%%FIG-4%%%%%%%%%%%%%%%%%%%%%%%%%%%%%%%%%%%%%%%%%%%%%

\begin{figure}[h]
\begin{center}
\begin{tabular}{c}
\resizebox{6cm}{6cm}{\includegraphics[angle=0]{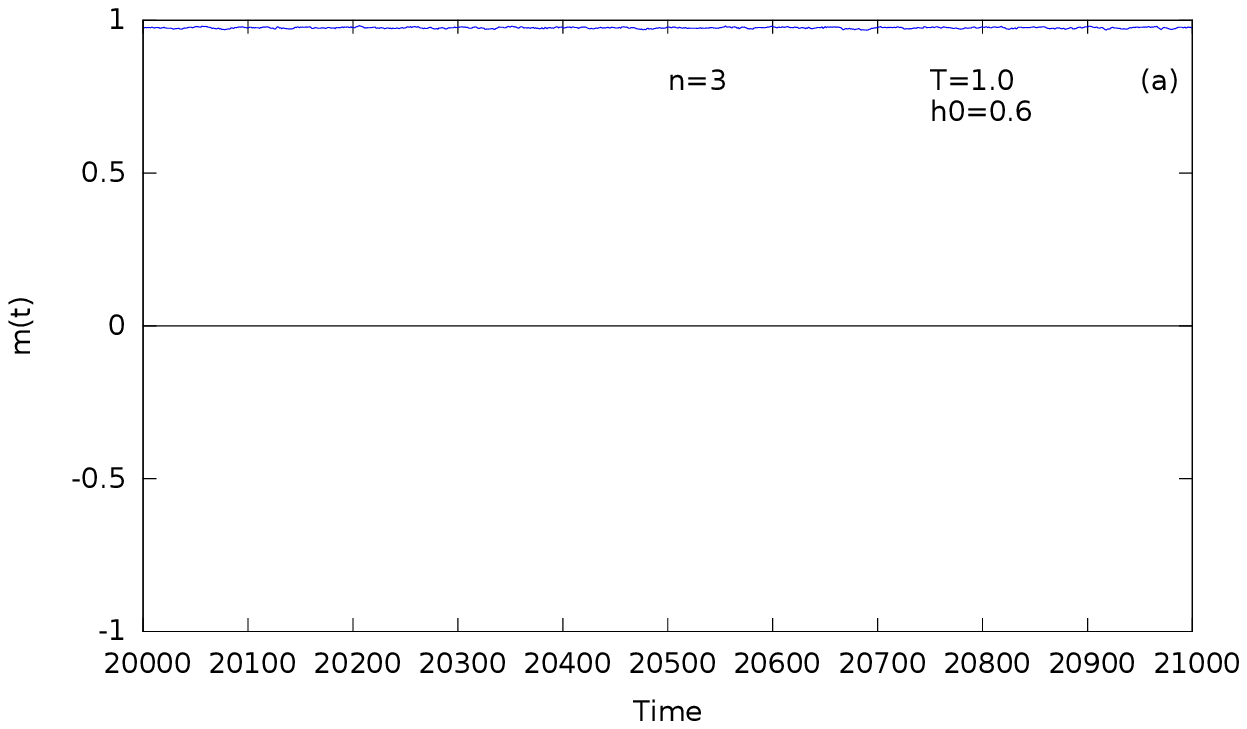}}
\resizebox{6cm}{6cm}{\includegraphics[angle=0]{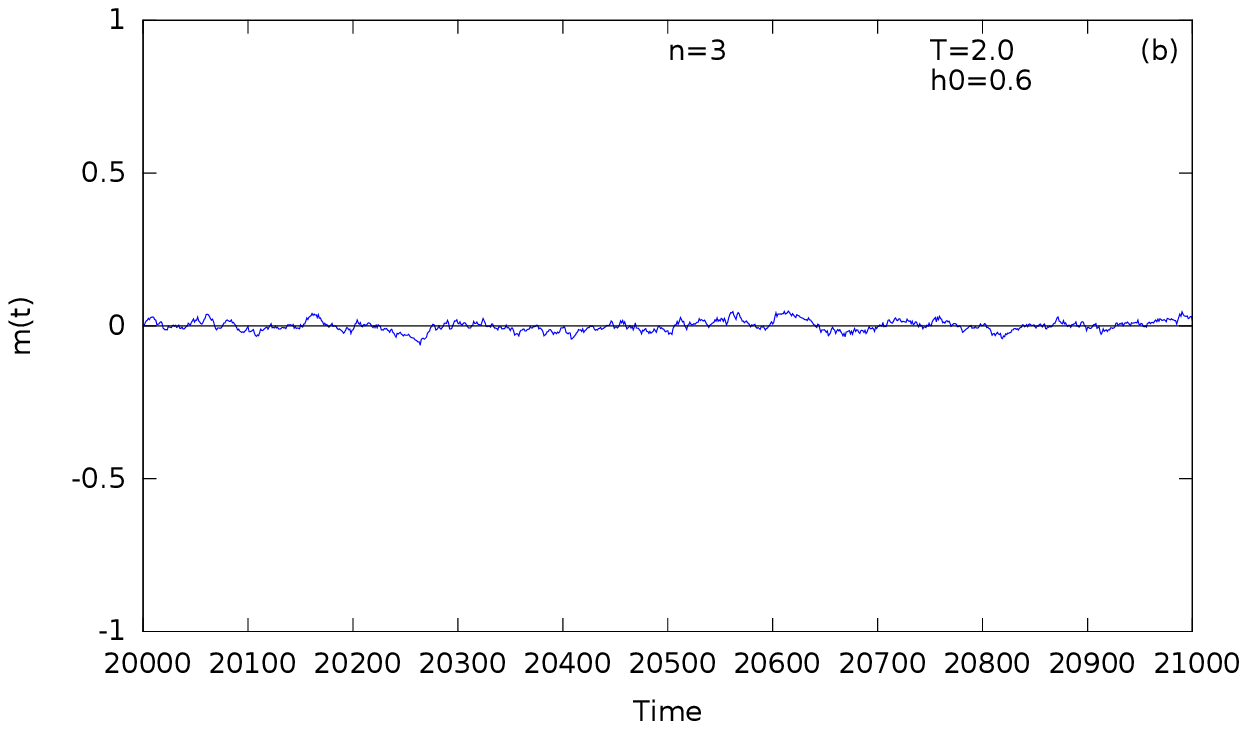}}
\\
\resizebox{6cm}{6cm}{\includegraphics[angle=0]{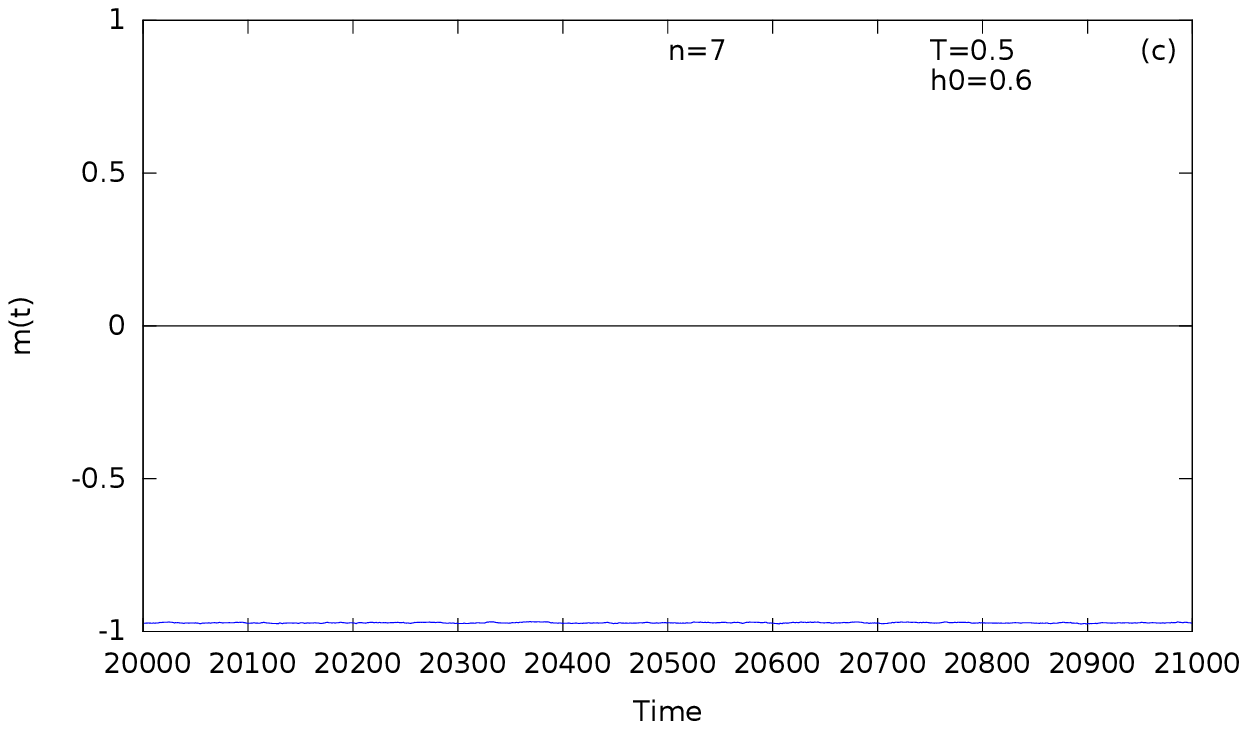}}
\resizebox{6cm}{6cm}{\includegraphics[angle=0]{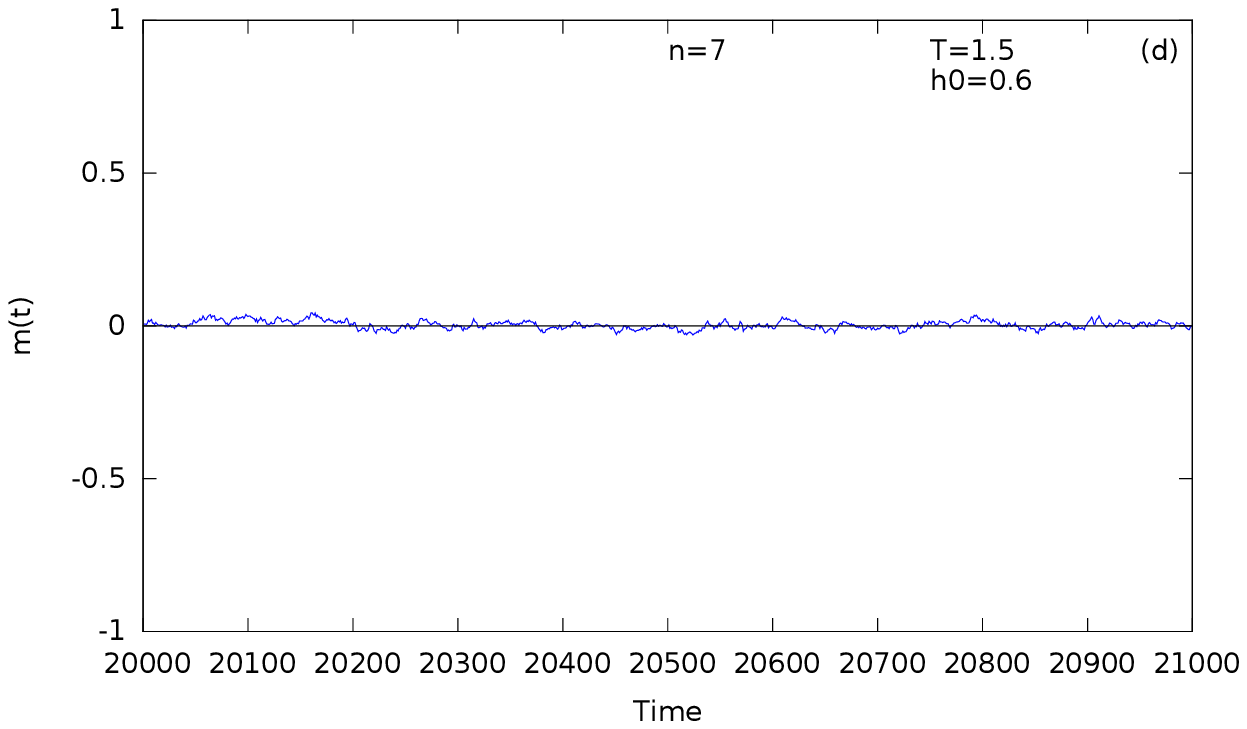}}
          \end{tabular}
\caption{\footnotesize{Variation of magnetisation with time at constant
 temperature ($T$) and field amplitude ($h_0$) for standing wave.
 Fig.4a \& fig.4b. represent {\it 3-state} spin whereas
 fig.4c. \& fig.4d. represent {\it 7-state} spin. Frequency $f=0.01$.}}

\end{center}
\end{figure}

%%%%%%%%%%%%%%%%%%%%%%%%%%%%%%%%%%%%%%%%%%%%%%%%%%%%%%%%%%%%%%%%%%%%%%%%%%%%%%%%%%
\newpage
%%%%%%%%%%%%%%%%FIG-5%%%%%%%%%%%%%%%%%%%%%%%%%%%%%%%%%%%%%%%%%%%%%%%%%%%%%%

\begin{figure}[h]
\begin{center}
\begin{tabular}{c}
\resizebox{6cm}{6cm}{\includegraphics[angle=0]{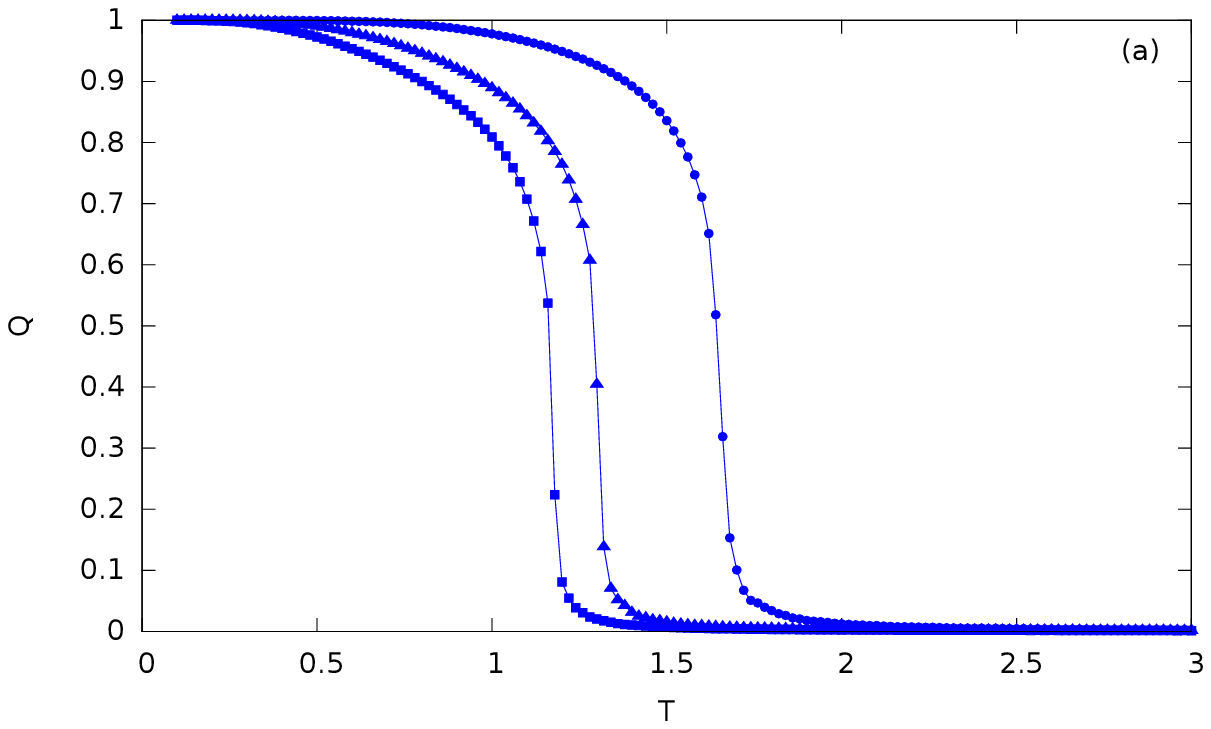}}
\resizebox{6cm}{6cm}{\includegraphics[angle=0]{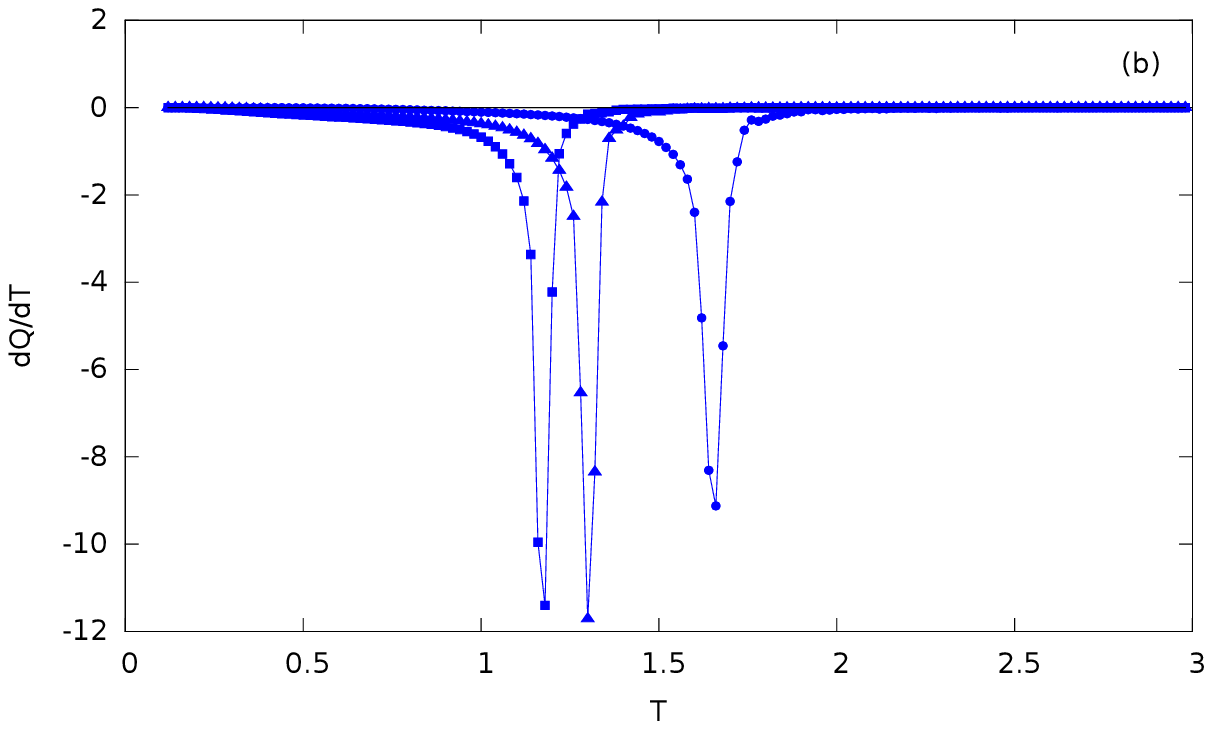}}
\\
\resizebox{6cm}{6cm}{\includegraphics[angle=0]{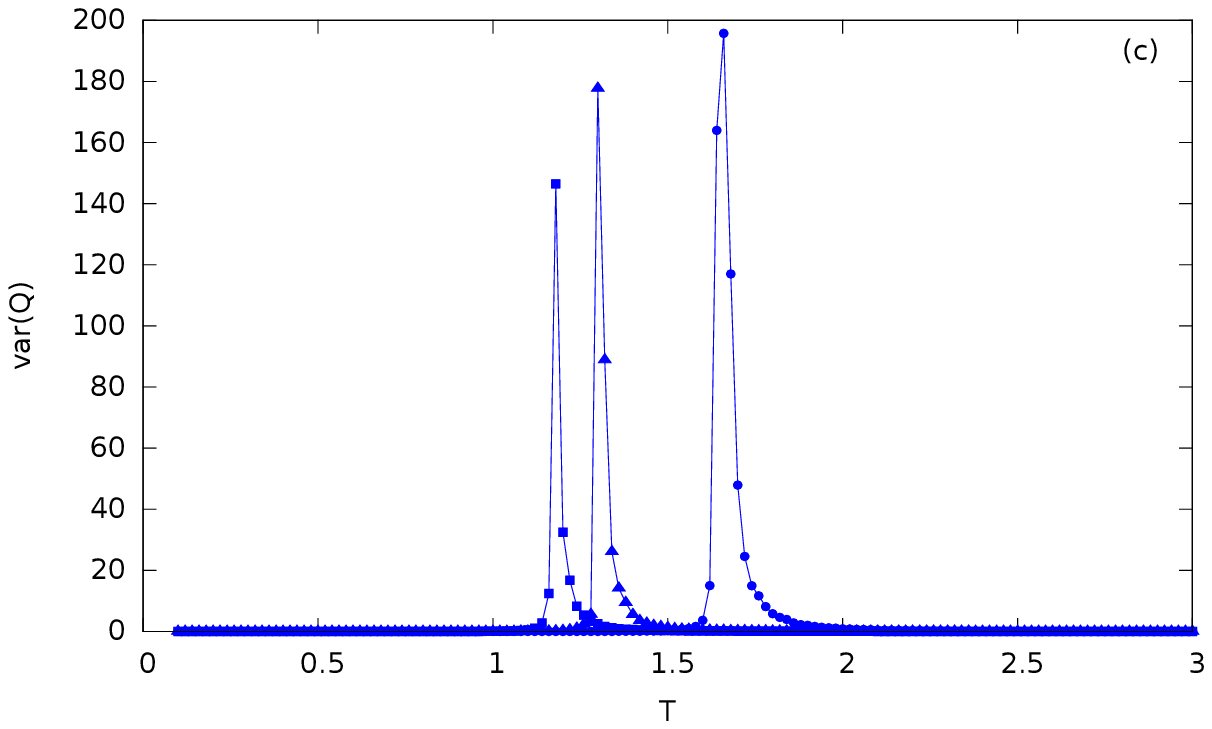}}
\resizebox{6cm}{6cm}{\includegraphics[angle=0]{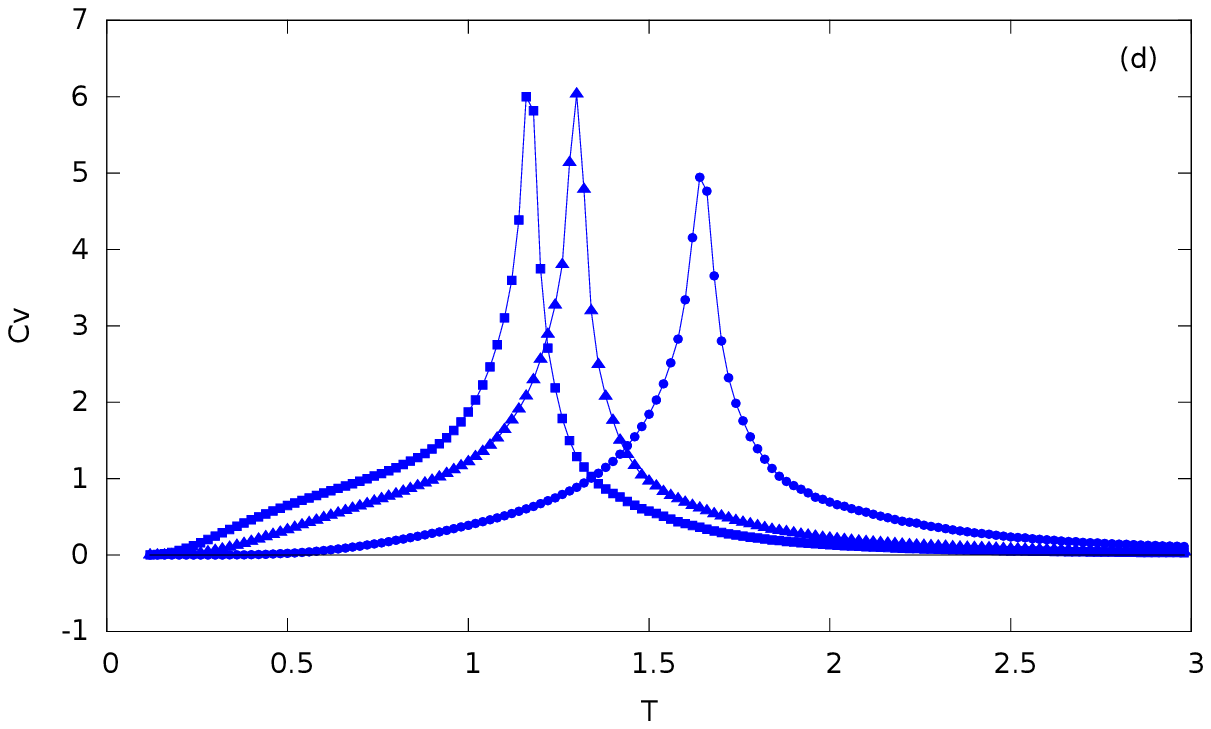}}
          \end{tabular}
\caption{\footnotesize{Temperature variation of (a) $Q$, (b) $\frac{dQ}{dT}$,
 (c) $V$, (d) $C_v$ for {\it 3-state} (circle), {\it 5-state} (uptriangle) and 
{\it 7-state} (square) spins for constant field amplitude ($h_0=0.2$)
 and frequency ($f=0.01$) of standing wave.}}

\end{center}
\end{figure}

%%%%%%%%%%%%%%%%%%%%%%%%%%%%%%%%%%%%%%%%%%%%%%%%%%%%%%%%%%%%%%%%%%%%%%%%%%%%%%%%%%
\newpage
%%%%%%%%%%%%%%%%FIG-6%%%%%%%%%%%%%%%%%%%%%%%%%%%%%%%%%%%%%%%%%

\begin{figure}[h]
\begin{center}
\begin{tabular}{c}
\resizebox{6cm}{6cm}{\includegraphics[angle=0]{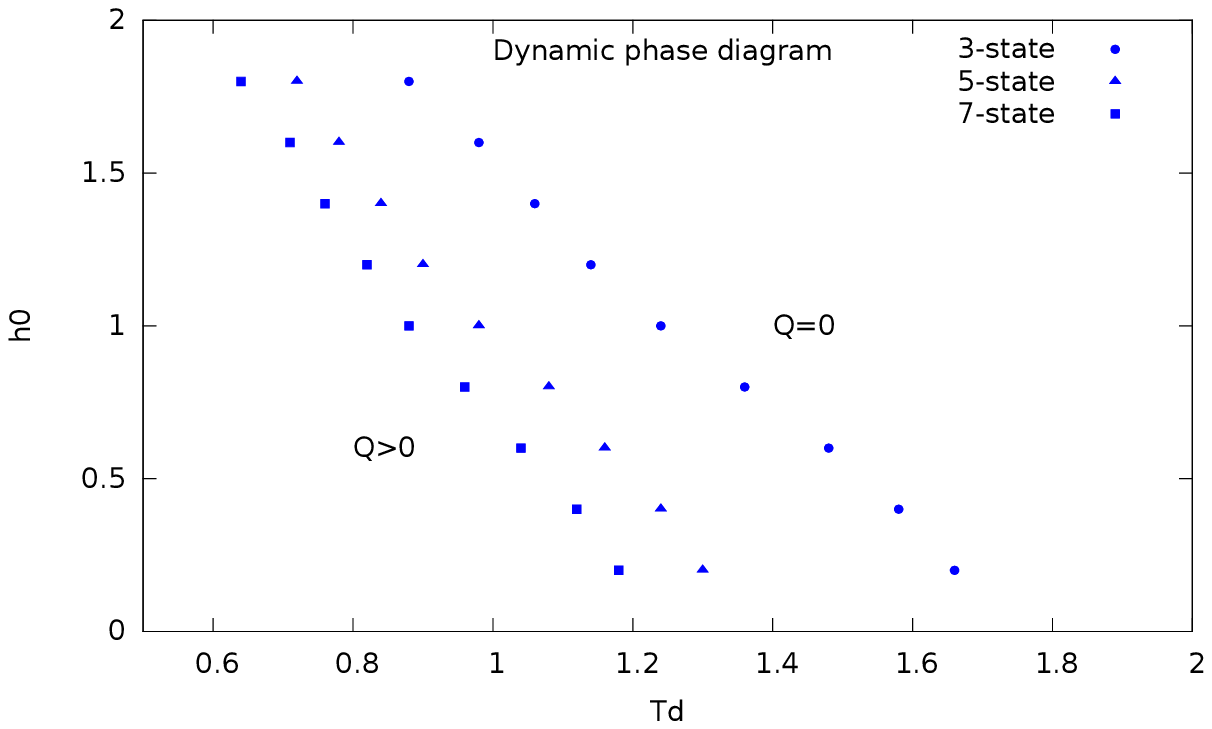}}
\\
          \end{tabular}
\caption{\footnotesize{Phase diagram in $T_d$-$h_0$ plane for {\it 3-state} (circle), {\it 5-state} (uptriangle) and 
{\it 7-state} (square) spins for standing wave.}}

\end{center}
\end{figure}

%%%%%%%%%%%%%%%%%%%%%%%%%%%%%%%%%%%%%%%%%%%%%%%%%%%%%%%%%%%%%%%%%%%%%%%%%%%%%%%%
\newpage
%%%%%%%%%%%%%%%%%%%%FIG-7%%%%%%%%%%%%%%%%%%%%%%%%%%%%%%%%%%%%%%%%

\begin{figure}[h]
\begin{center}
\begin{tabular}{c}
\resizebox{6cm}{6cm}{\includegraphics[angle=0]{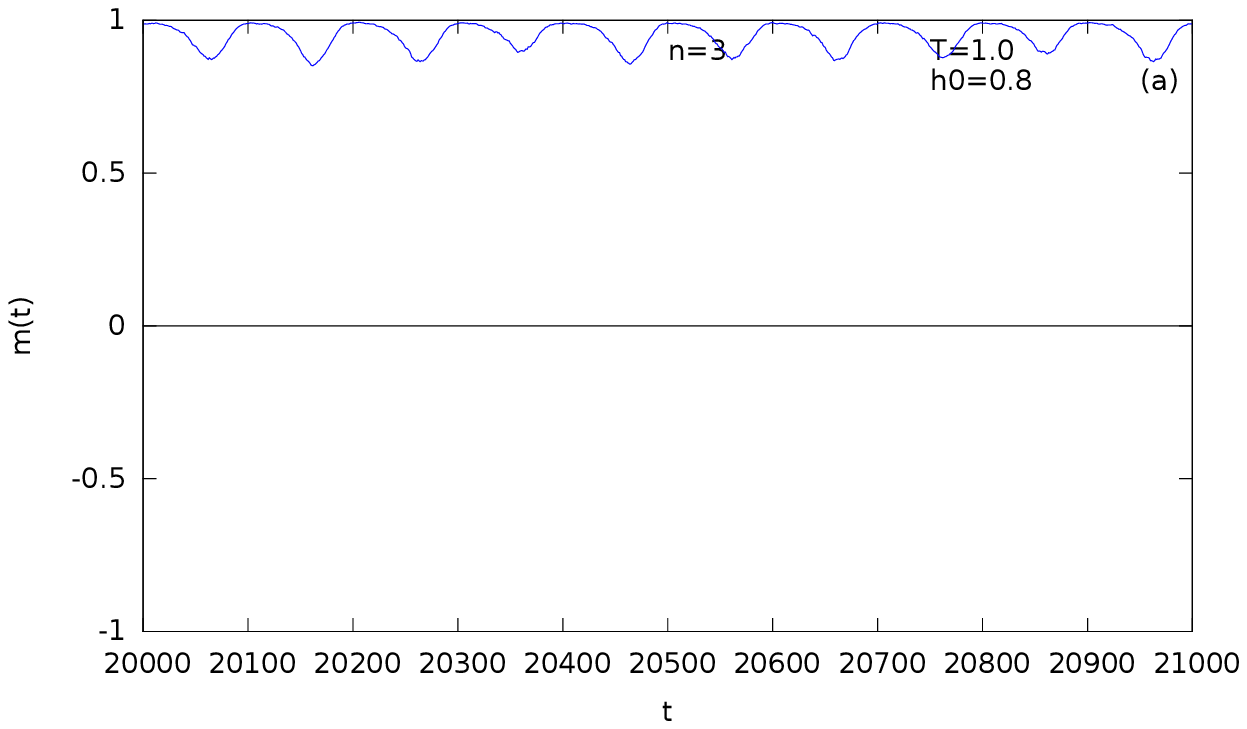}}
\resizebox{6cm}{6cm}{\includegraphics[angle=0]{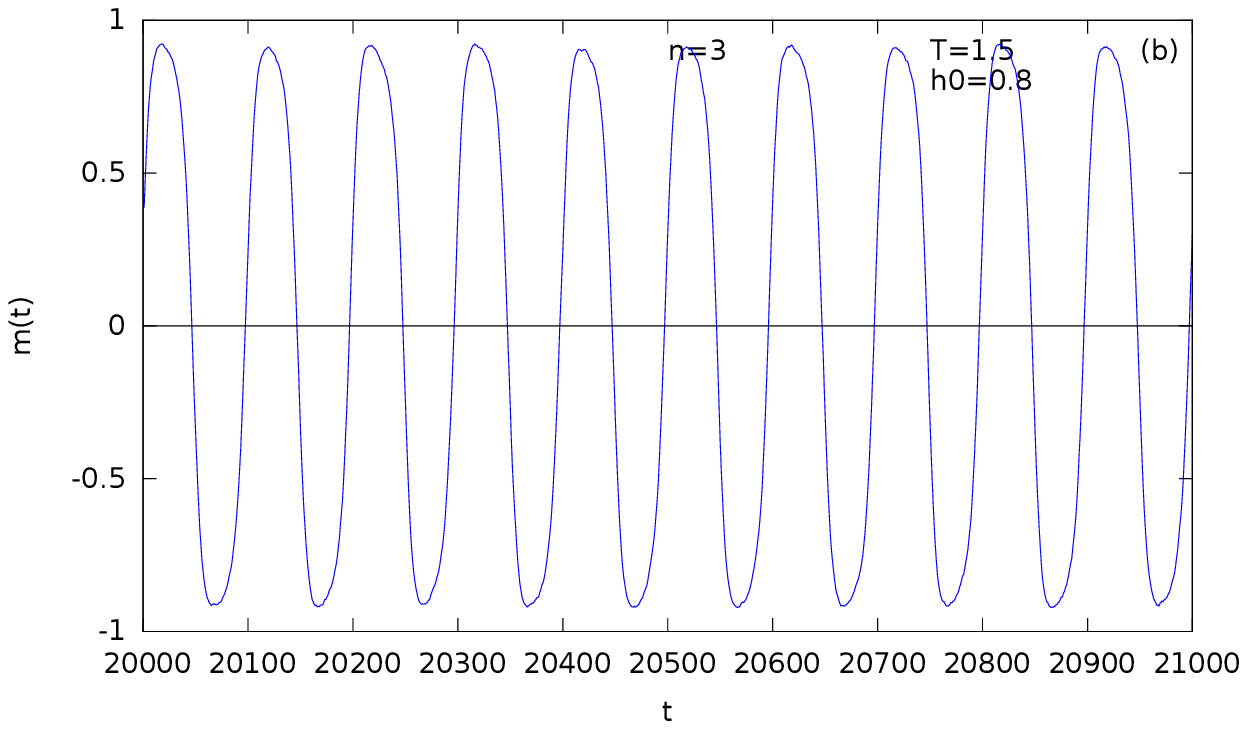}}
\\
\resizebox{6cm}{6cm}{\includegraphics[angle=0]{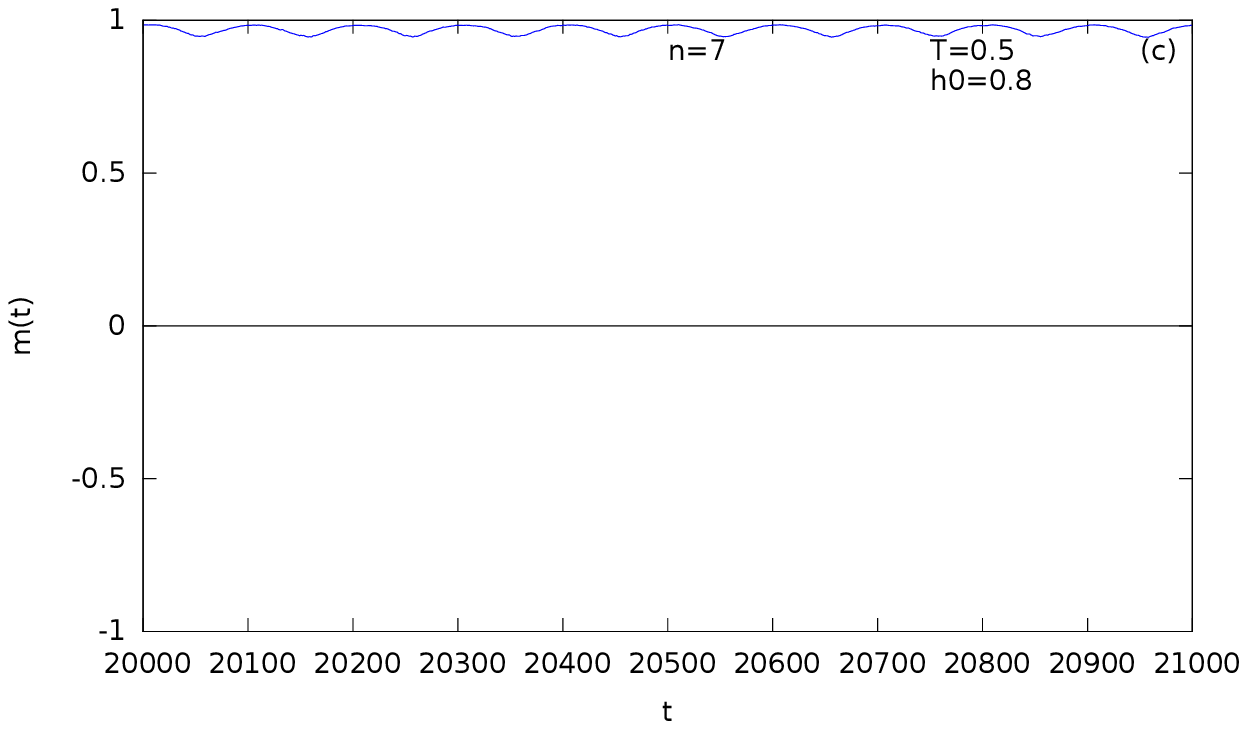}}
\resizebox{6cm}{6cm}{\includegraphics[angle=0]{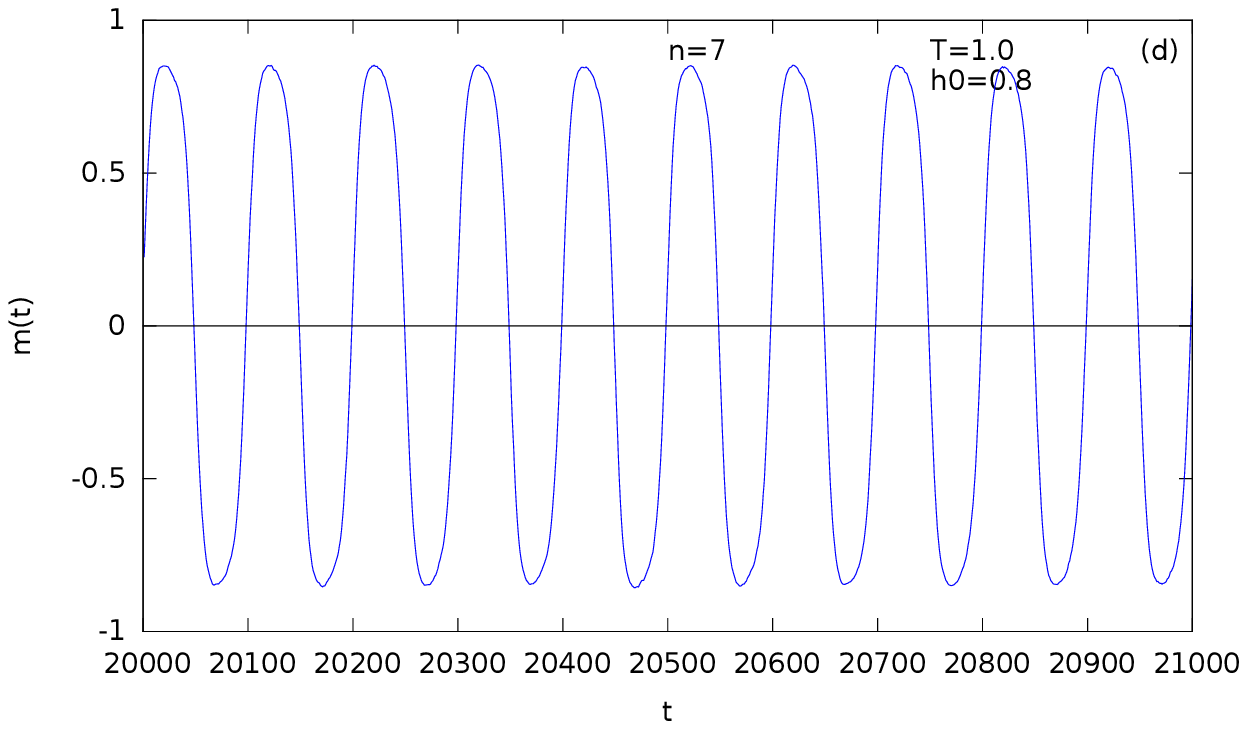}}
          \end{tabular}
\caption{\footnotesize{Variation of magnetisation with time at constant
 temperature ($T$) and field amplitude ($h_0$) for uniformly varying field.
 Fig.7a \& fig.7b. represent {\it 3-state} spin whereas
 fig.7c. \& fig.7d. represent {\it 7-state} spin. Frequency $f=0.01$.}}

\end{center}
\end{figure}

%%%%%%%%%%%%%%%%%%%%%%%%%%%%%%%%%%%%%%%%%%%%%%%%%%%%%%%%%%%%%%%%%%%%%%%%%%%%%%%%%%
\newpage
%%%%%%%%%%%%%%%%FIG-8%%%%%%%%%%%%%%%%%%%%%%%%%%%%%%%%%%%%%%%%%%%%%%%%%%

\begin{figure}[h]
\begin{center}
\begin{tabular}{c}
\resizebox{6cm}{6cm}{\includegraphics[angle=0]{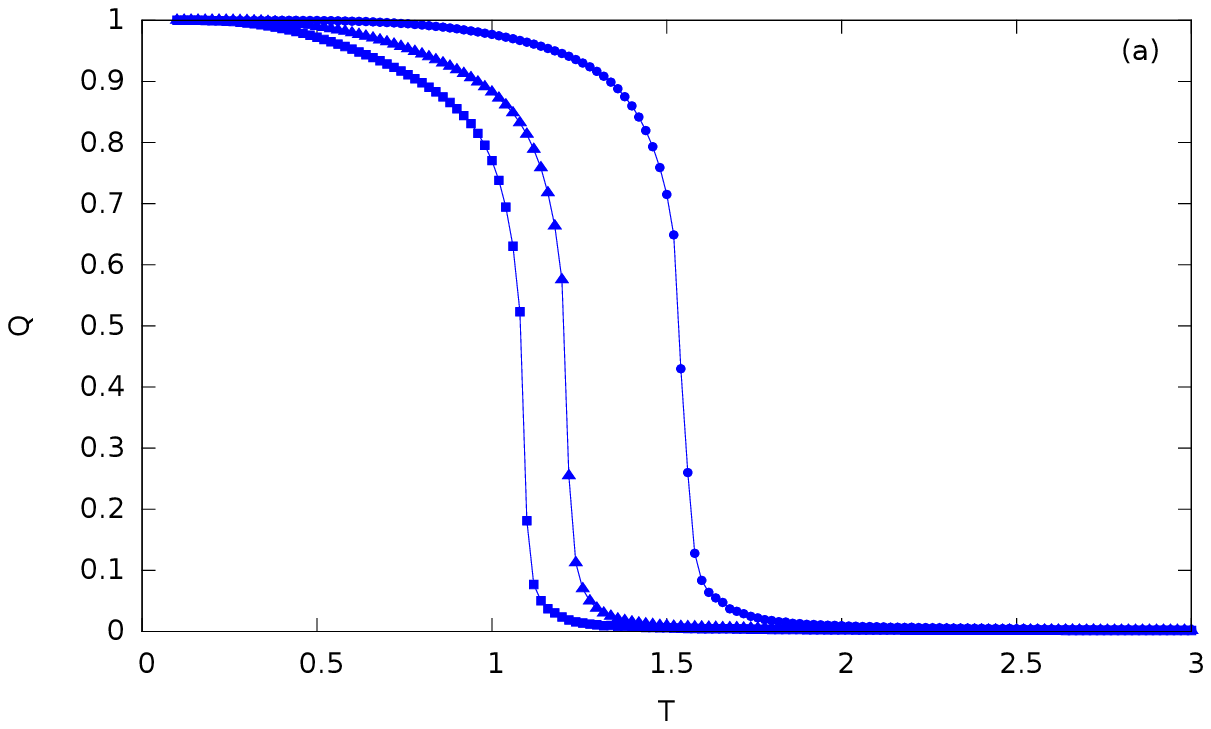}}
\resizebox{6cm}{6cm}{\includegraphics[angle=0]{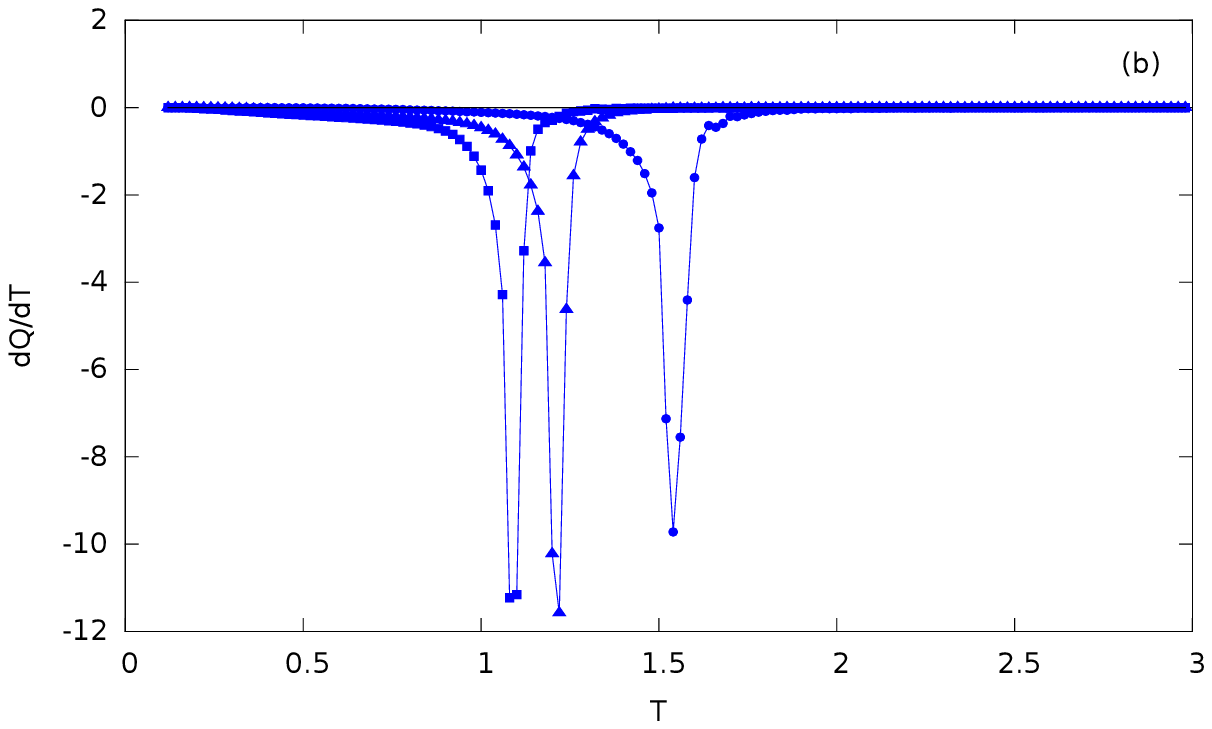}}
\\
\resizebox{6cm}{6cm}{\includegraphics[angle=0]{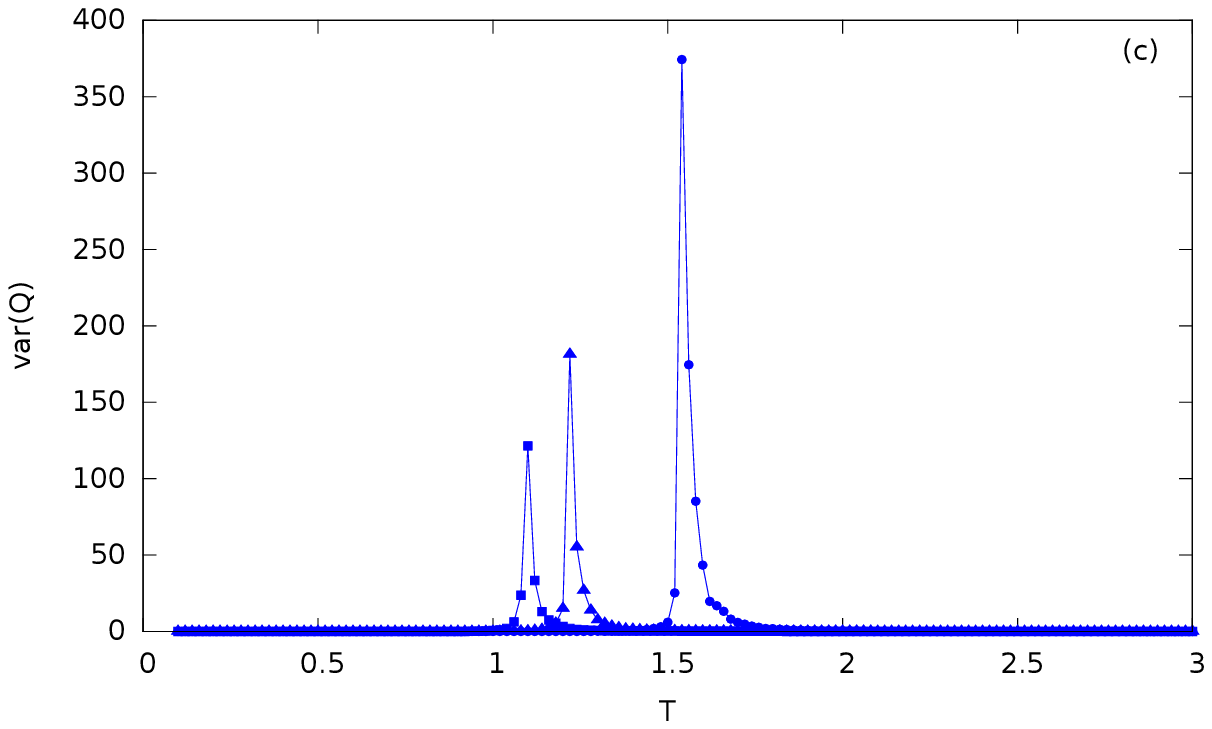}}
\resizebox{6cm}{6cm}{\includegraphics[angle=0]{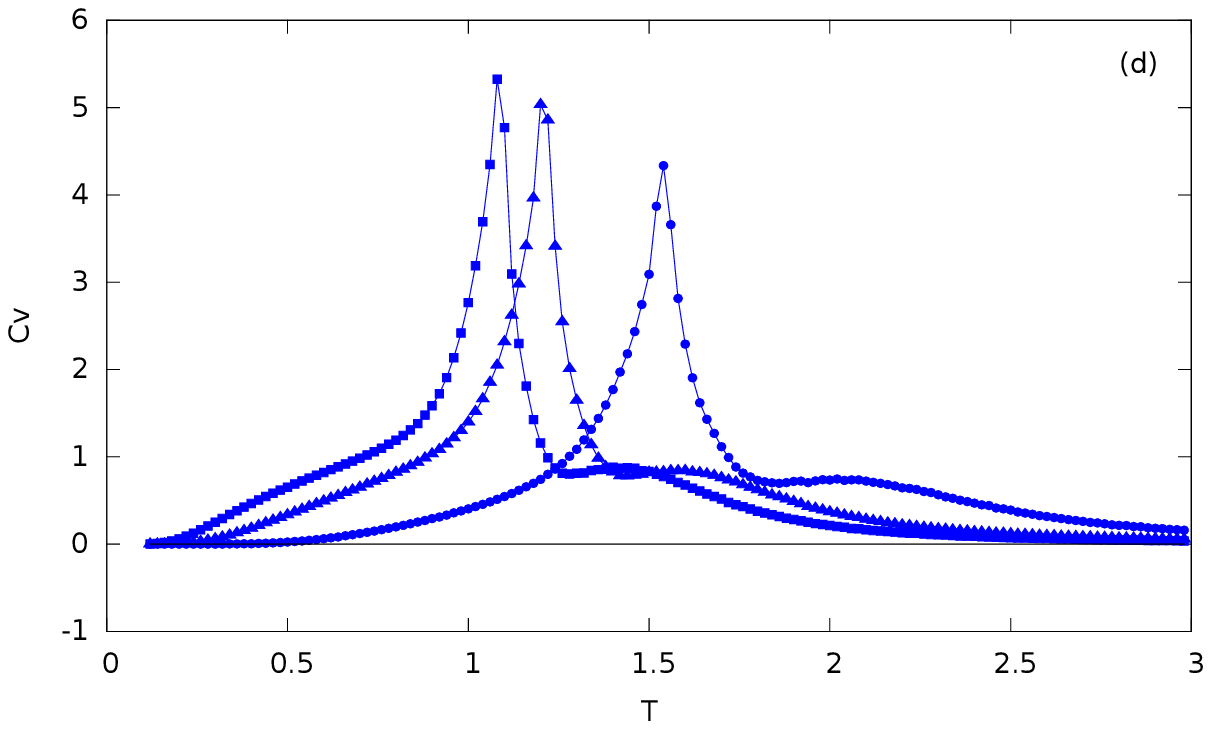}}
          \end{tabular}
\caption{\footnotesize{Temperature variation of (a) $Q$, (b) $\frac{dQ}{dT}$,
 (c) $V$, (d) $C_v$ for {\it 3-state} (circle), {\it 5-state} (uptriangle) and 
{\it 7-state} (square) spins for constant field amplitude ($h_0=0.3$)
 and frequency ($f=0.01$) of uniformly varying field.}}

\end{center}
\end{figure}

%%%%%%%%%%%%%%%%%%%%%%%%%%%%%%%%%%%%%%%%%%%%%%%%%%%%%%%%%%%%%%%%%%%%%%%%%%%%%%%%%%
\newpage
%%%%%%%%%%%%%%%%%%%%FIG-9%%%%%%%%%%%%%%%%%%%%%%%%%%%%%%%%%%%%%%%

\begin{figure}[h]
\begin{center}
\begin{tabular}{c}
\resizebox{6cm}{6cm}{\includegraphics[angle=0]{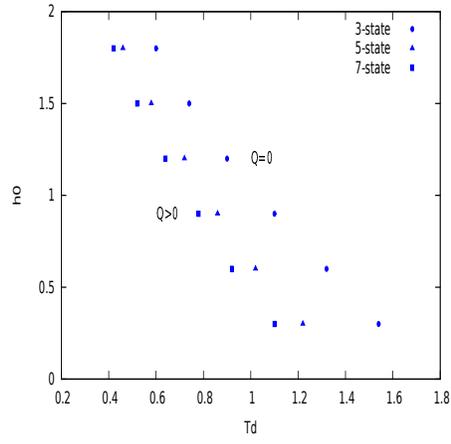}}
\\
          \end{tabular}
\caption{\footnotesize{Phase diagram in $T_d$-$h_0$ plane for {\it 3-state} (circle), {\it 5-state} (uptriangle) and 
{\it 7-state} (square) spins for uniformly varying field.}}

\end{center}
\end{figure}

%%%%%%%%%%%%%%%%%%%%%%%%%%%%%%%%%%%%%%%%%%%%%%%%%%%%%%%%%%%%%%%%%%%%%%%%%%%%%%%%%%
%%%%%%%%%%%%%%%%%%%%%%%%FIG-10%%%%%%%%%%%%%%%%%%%%%%%%%%%%%%%%

\begin{figure}[h]
\begin{center}
\begin{tabular}{c}
{\includegraphics[angle=0]{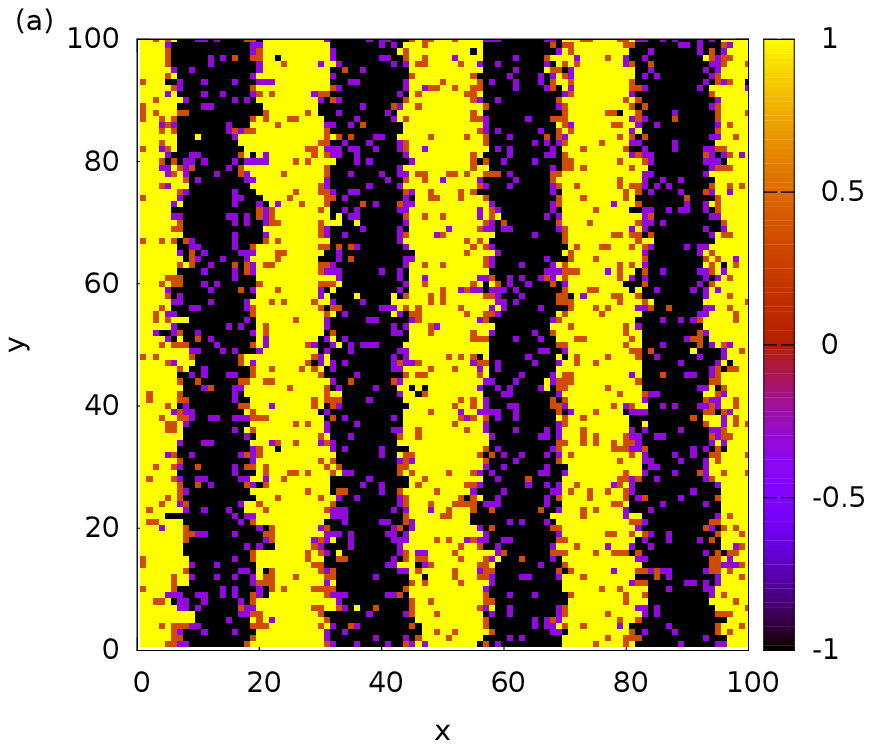}}
\\
{\includegraphics[angle=0]{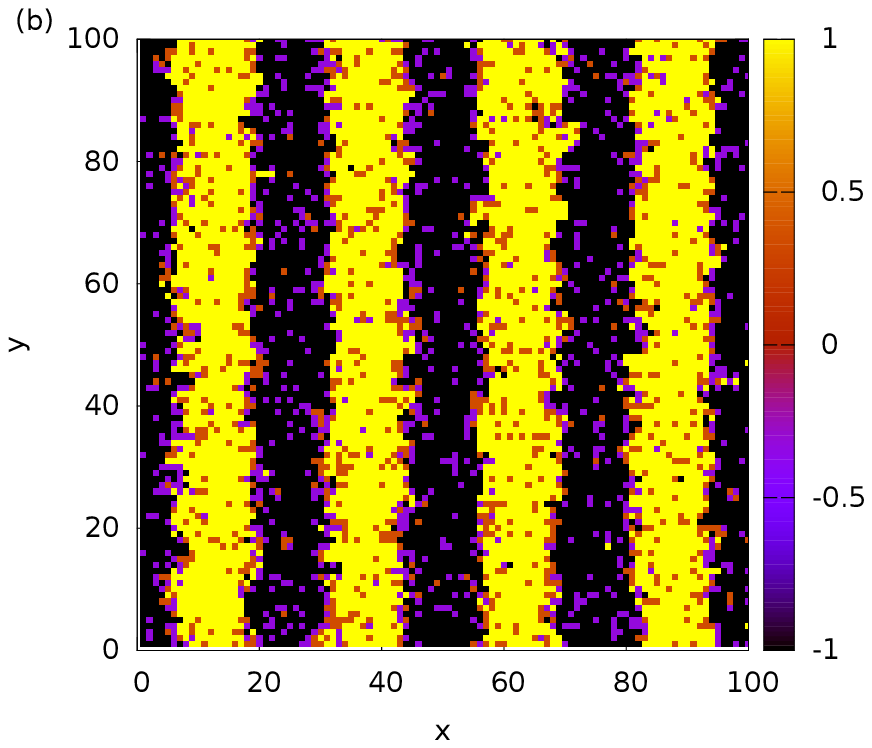}}
          \end{tabular}
\caption{\footnotesize{Coherent propagation of spin-wave is shown at two different 
times for propagating magnetic wave:
 (a) at $t=4018$ and (b) at $t=4067$. Here $n=4$, $h_0=1.0$ \& $T=1.2$.}}

\end{center}
\end{figure}

%%%%%%%%%%%%%%%%%%%%%%%%%%%%%%%%%%%%%%%%%%%%%%%%%%%%%%%%%%%%%%%%%%%%%%%%%%%%%%%%%%
%%%%%%%%%%%%%%%%%%%%%%%%%%%%%FIG-11%%%%%%%%%%%%%%%%%%%%%%%%%%%%

\begin{figure}[h]
\begin{center}
\begin{tabular}{c}
\resizebox{7cm}{6cm}{\includegraphics[angle=0]{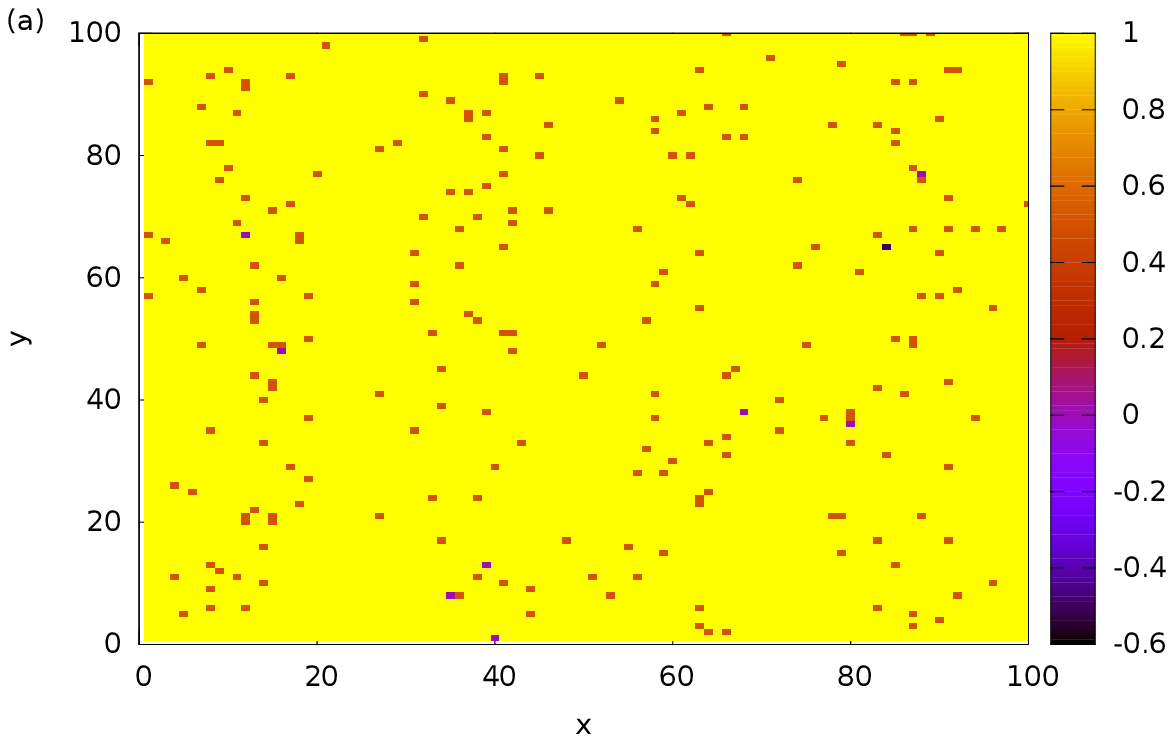}}
\resizebox{7cm}{6cm}{\includegraphics[angle=0]{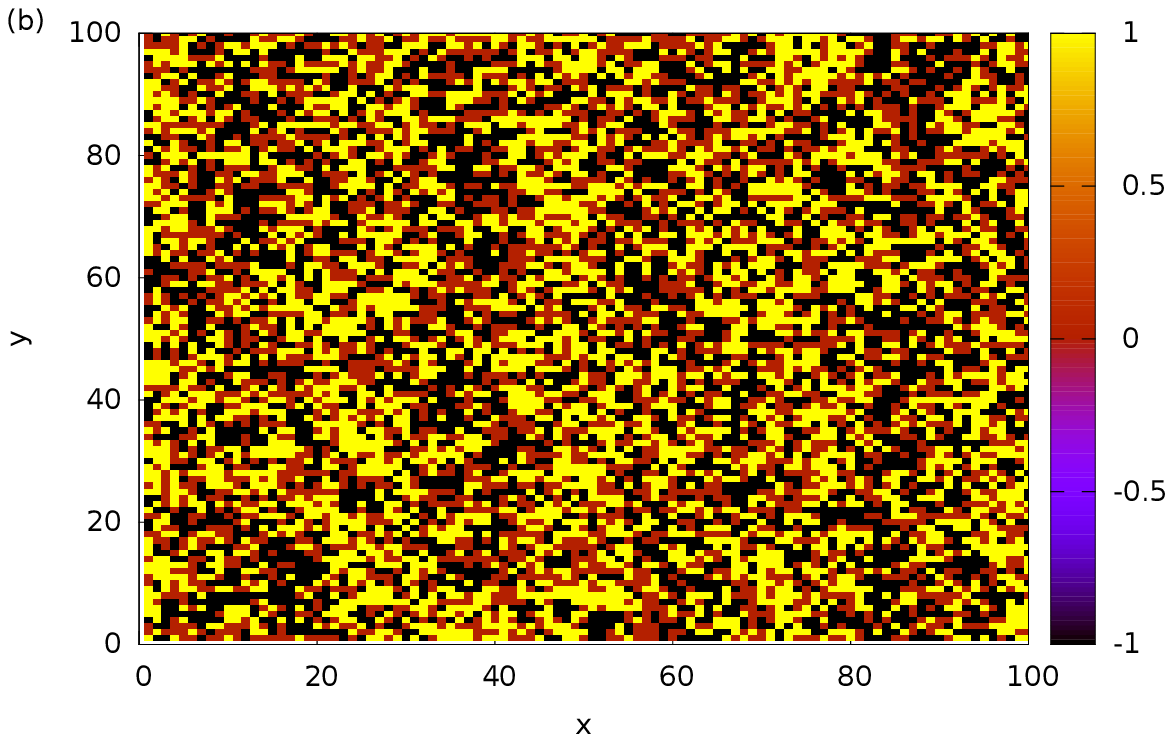}}
\\
\resizebox{7cm}{6cm}{\includegraphics[angle=0]{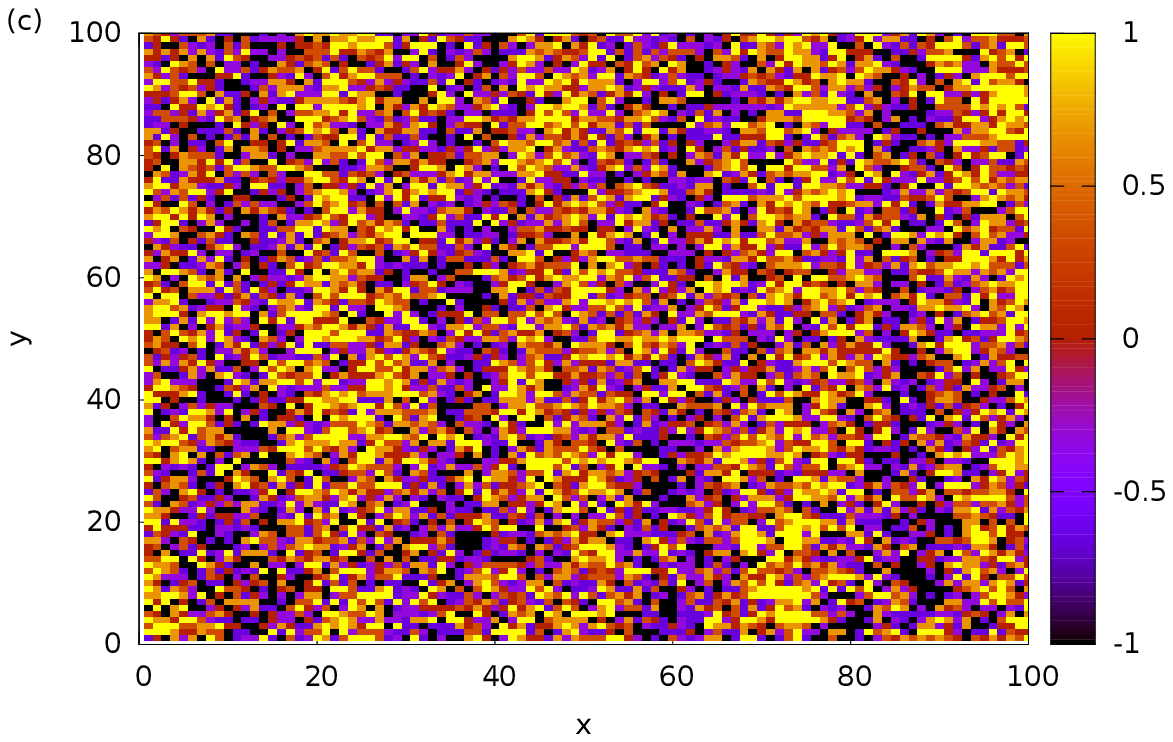}}
\resizebox{7cm}{6cm}{\includegraphics[angle=0]{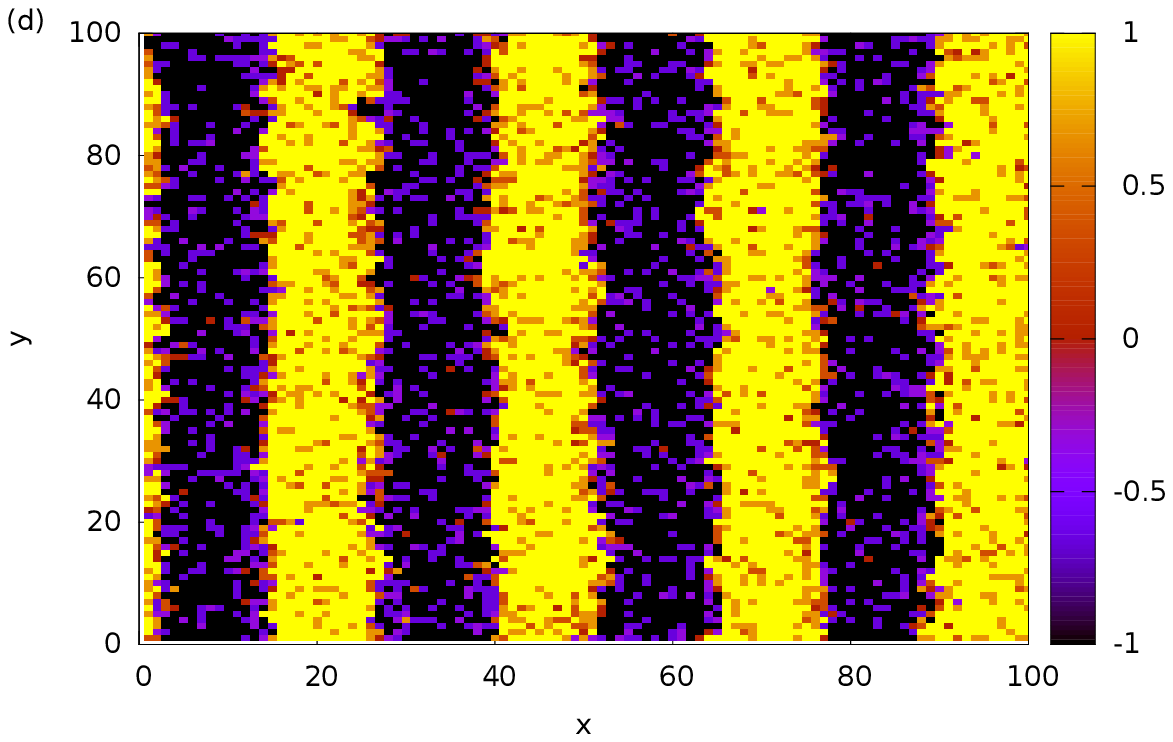}}
          \end{tabular}
\caption{\footnotesize{Lattice morphology (value of $S^z(x,y,t)$) at 
time $t=4000$ for different values of $n$, $h_0$ and $T$ for propagating magnetic wave: 
(a) $n=5$, $h_0=0.8$ \& $T=0.5$, (b) $n=3$, $h_0=0.2$ \& $T=2.8$, 
(c) $n=7$, $h_0=0.3$ \& $T=1.8$ and (d) $n=7$, $h_0=1.2$ \& $T=0.8$.}}

\end{center}
\end{figure}

%%%%%%%%%%%%%%%%%%%%%%%%%%%%%%%%%%%%%%%%%%%%%%%%%%%%%%%%%%%%%%%%%%%%%%%%%%%%%%%%%%
%%%%%%%%%%%%%%%%%%%%%%%%%%%%%%%%%%FIG-12%%%%%%%%%%%%%%%%%%%%%%%%

\begin{figure}[h]
\begin{center}
\begin{tabular}{c}
{\includegraphics[angle=0]{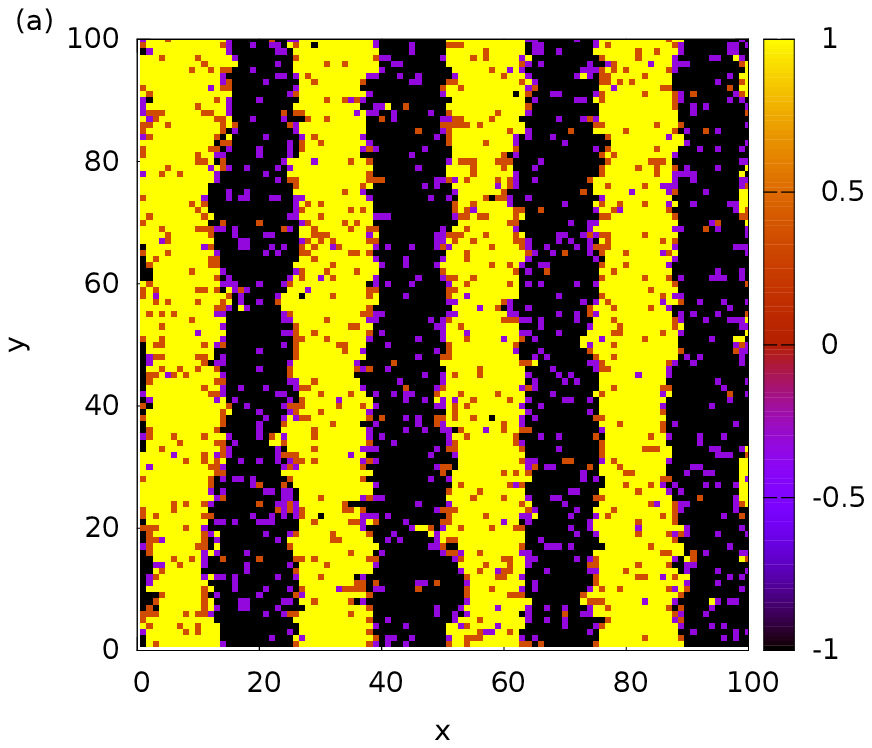}}
\\
{\includegraphics[angle=0]{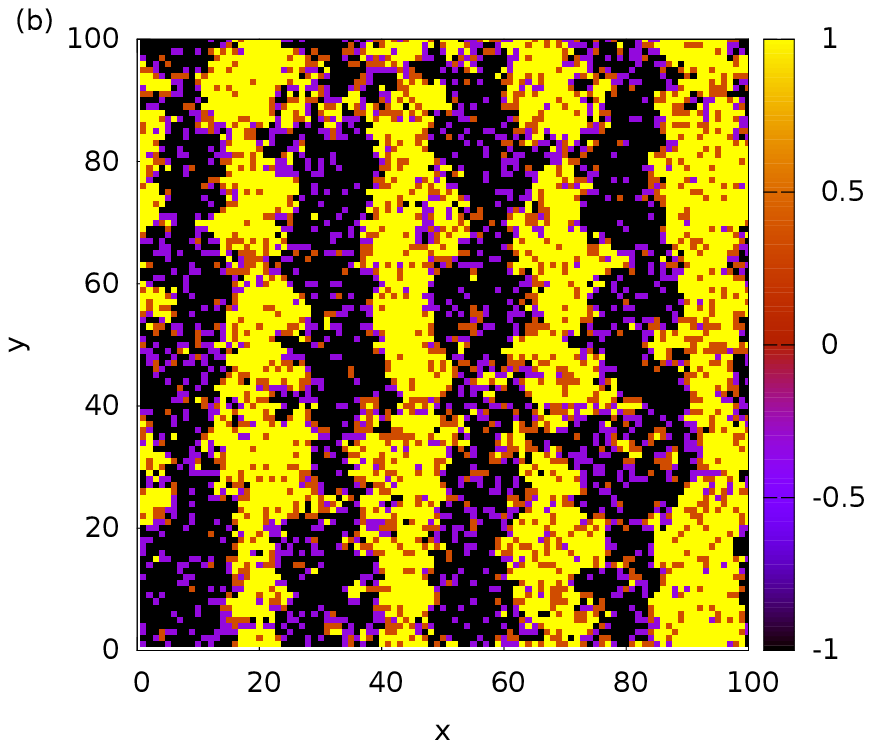}}
          \end{tabular}
\caption{\footnotesize{Morphologies of standing wave dynamical modes 
(non-propagating) for standing magnetic wave are shown at two different 
times: (a) at $t=4040$ and (b) at $t=4067$. Here $n=4$, $h_0=1.2$ \& $T=1.2$.}}

\end{center}
\end{figure}

%%%%%%%%%%%%%%%%%%%%%%%%%%%%%%%%%%%%%%%%%%%%%%%%%%%%%%%%%%%%%%%%%%%%%%%%%%%%%%%%%%
%%%%%%%%%%%%%%%%%%%%%%%%%%%%%%%%%%%%%%%FIG-13%%%%%%%%%%%%%%%

\begin{figure}[h]
\begin{center}
\begin{tabular}{c}
\resizebox{7cm}{6cm}{\includegraphics[angle=0]{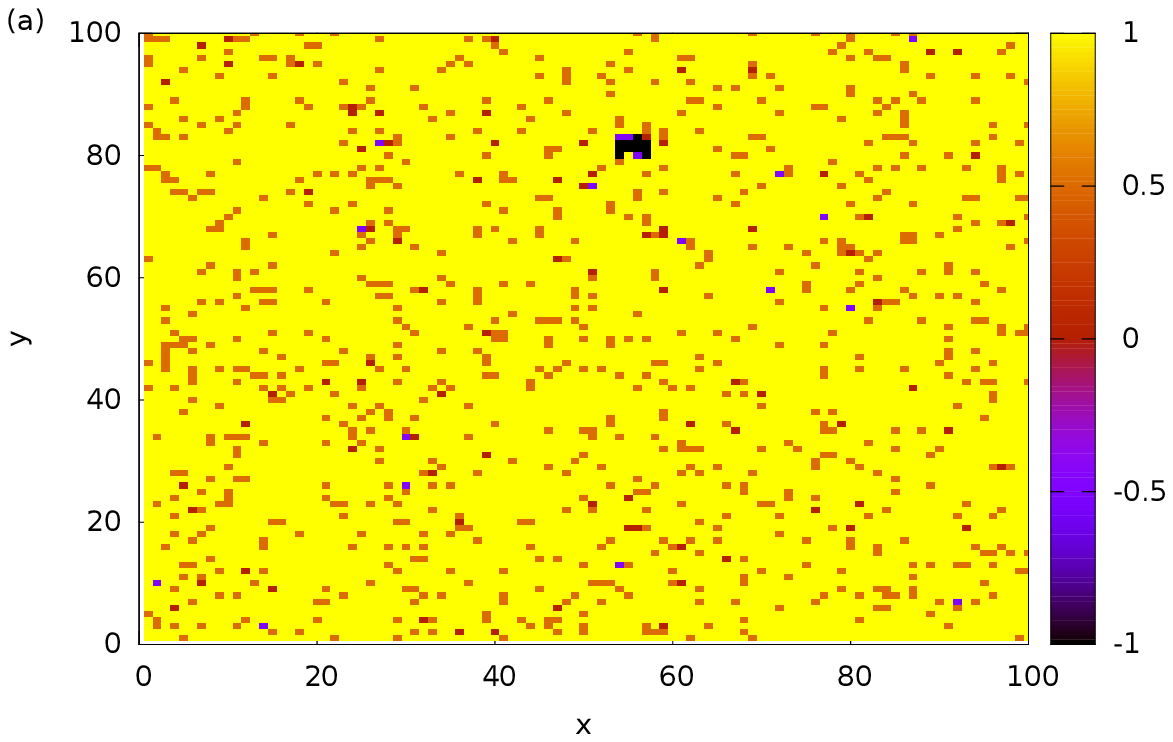}}
\resizebox{7cm}{6cm}{\includegraphics[angle=0]{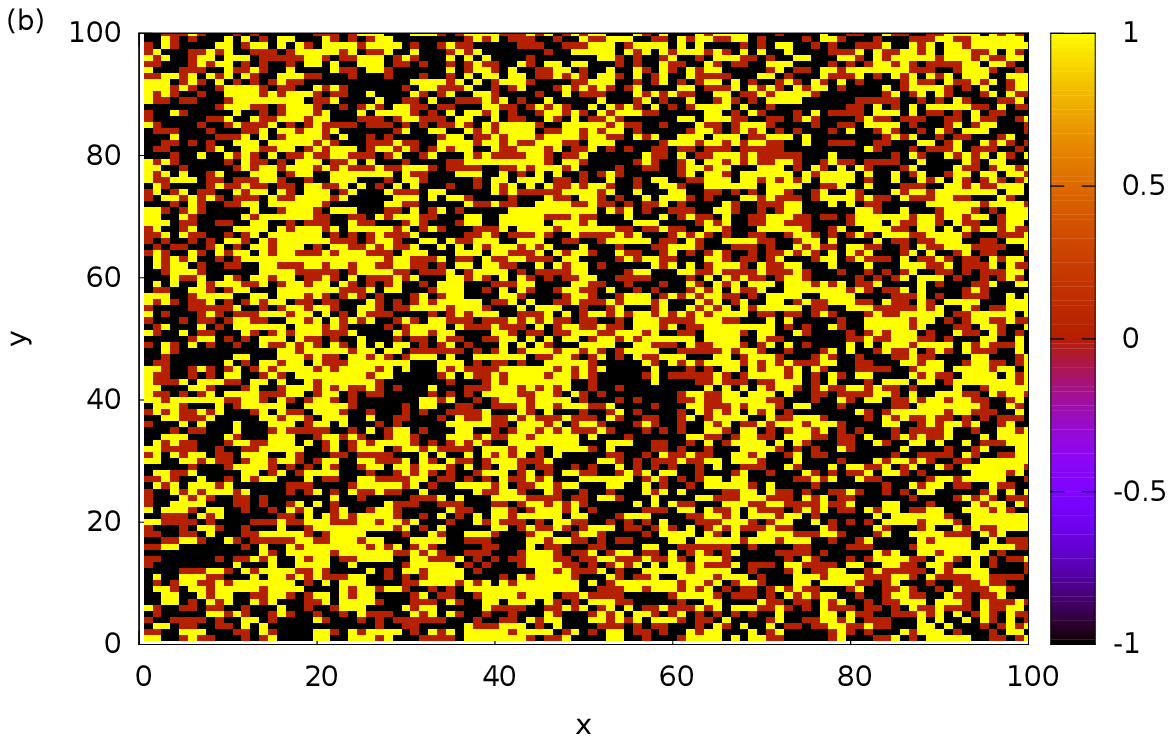}}
\\
\resizebox{7cm}{6cm}{\includegraphics[angle=0]{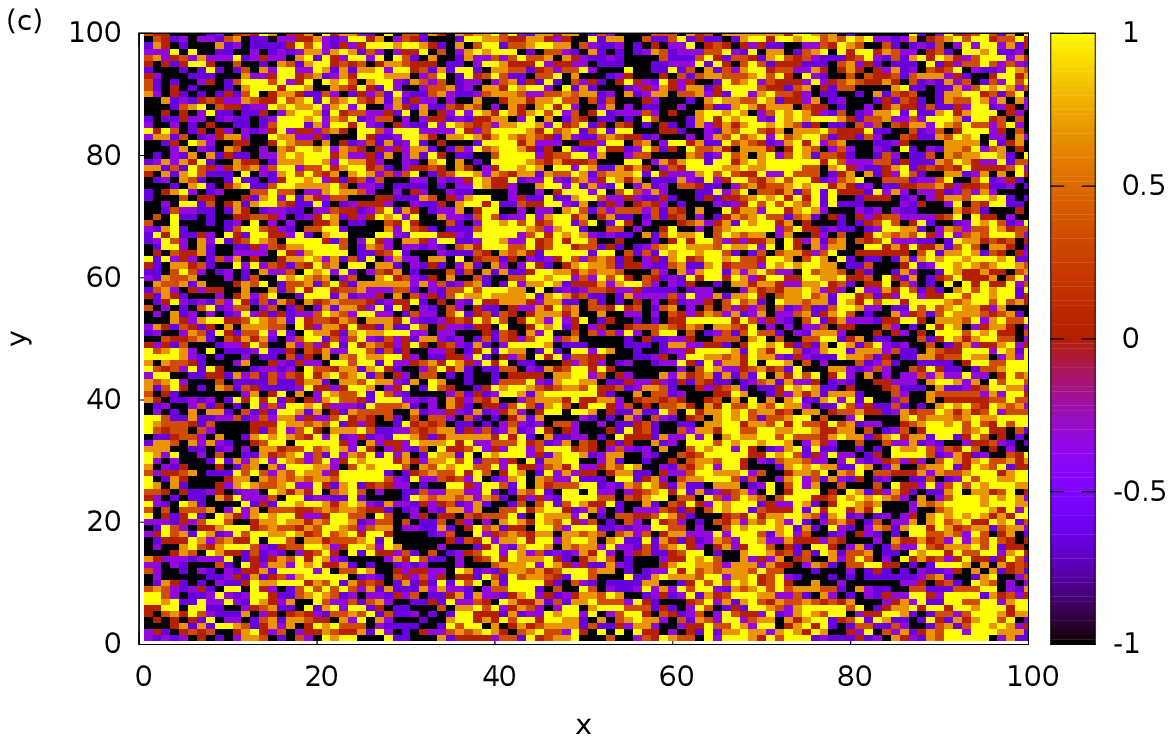}}
\resizebox{7cm}{6cm}{\includegraphics[angle=0]{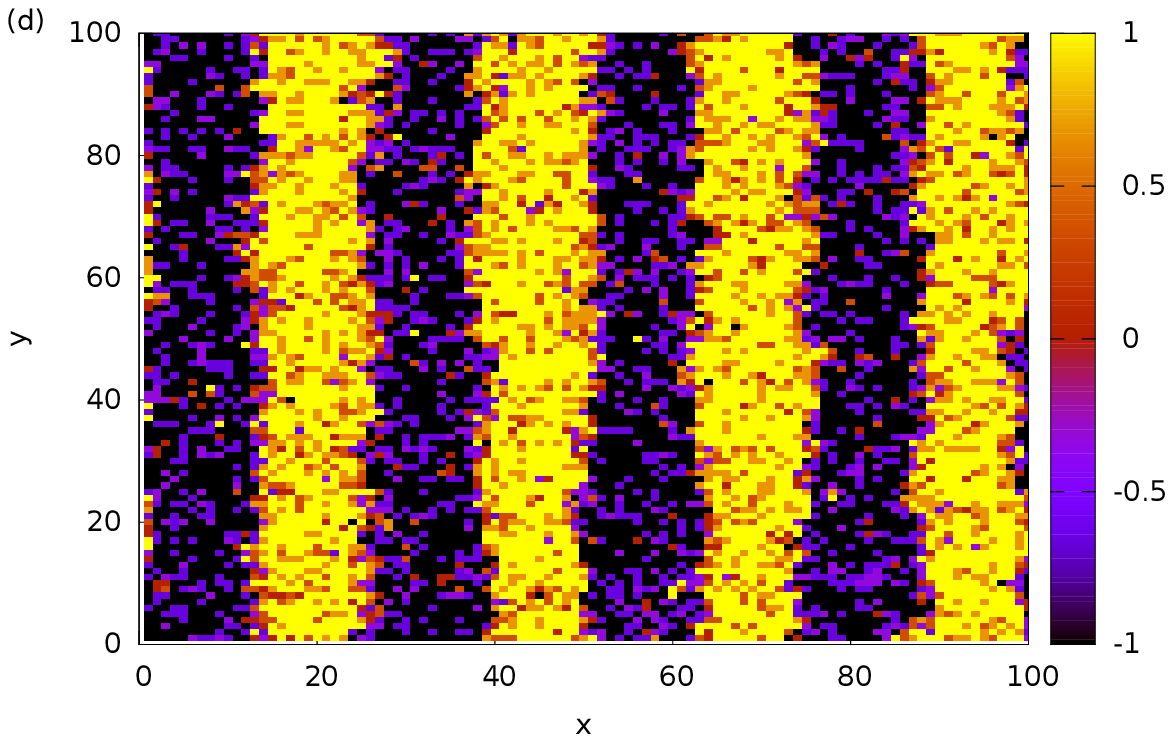}}
          \end{tabular}
\caption{\footnotesize{Lattice morphology (value of $S^z(x,y,t)$) at 
time $t=4000$ for different values of $n$, $h_0$ and $T$ for standing magnetic wave: 
(a) $n=5$, $h_0=0.8$ \& $T=0.8$, (b) $n=3$, $h_0=0.2$ \& $T=2.2$, 
(c) $n=7$, $h_0=0.3$ \& $T=1.5$, \& (d) $n=7$, $h_0=1.2$ \& $T=1.0$.}}

\end{center}
\end{figure}

%%%%%%%%%%%%%%%%%%%%%%%%%%%%%%%%%%%%%%%%%%%%%%%%%%%%%%%%%%%%%%%%%%%%%%%%%%%%%%%%%%
%%%%%%%%%%%%%%%%%%%%%%%%%%%%%%%%%%%%%%%%%%%%FIG-14%%%%%%%%%%%%%%

\begin{figure}[h]
\begin{center}
\begin{tabular}{c}
{\includegraphics[angle=0]{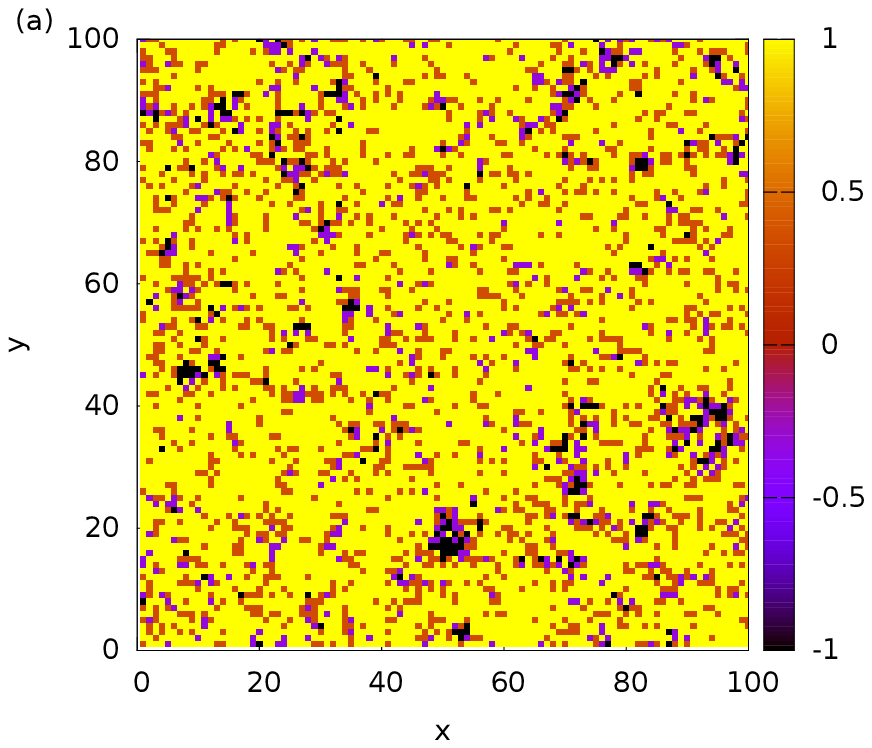}}
\\
{\includegraphics[angle=0]{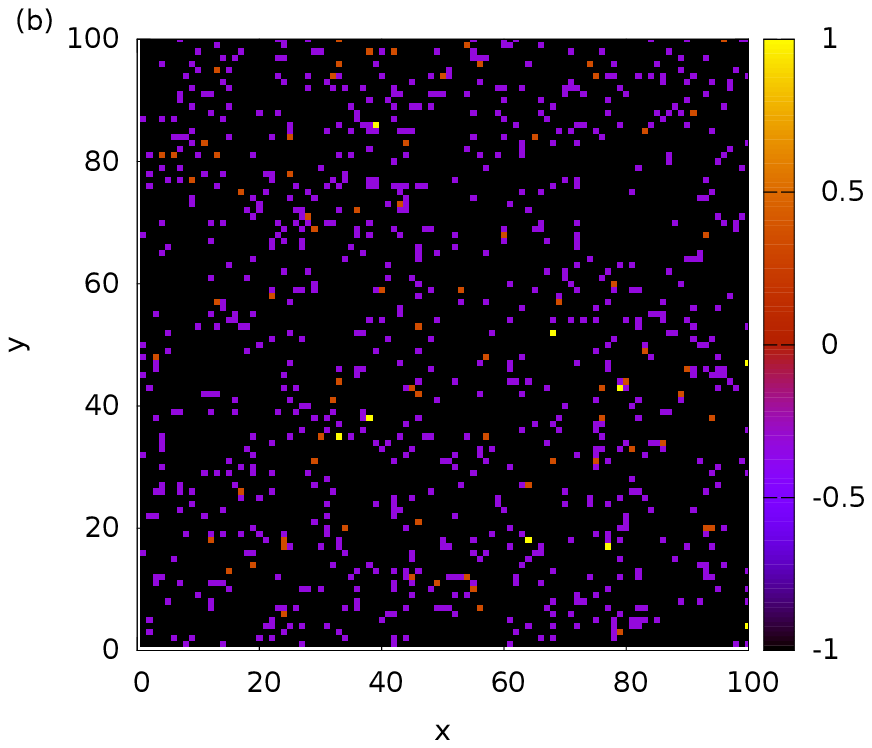}}
          \end{tabular}
\caption{\footnotesize{Oscillation of spins shown at different times ($t=4000$ \& $t=4067$)
 in symmetric phase for uniformly oscillating magnetic field.
Here $n=4$, $h_0=1.2$ \& $T=1.2$.}}

\end{center}
\end{figure}

%%%%%%%%%%%%%%%%%%%%%%%%%%%%%%%%%%%%%%%%%%%%%%%%%%%%%%%%%%%%%%%%%%%%%%%%%%%%%%%%%%
%%%%%%%%%%%%%%%%%%%%%%%%%%%%%%%%%%%%%%%%%%%%%%%%%FIG-15%%%%%%%%%%

\begin{figure}[h]
\begin{center}
\begin{tabular}{c}
\resizebox{7cm}{6cm}{\includegraphics[angle=0]{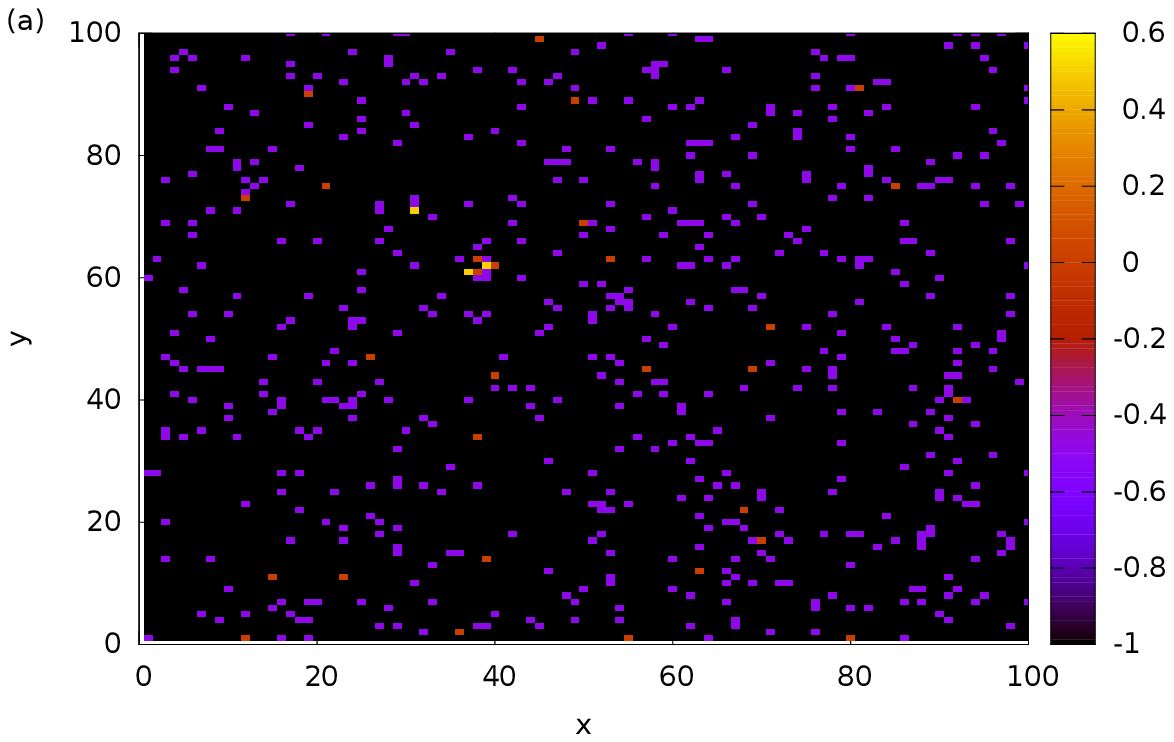}}
\resizebox{7cm}{6cm}{\includegraphics[angle=0]{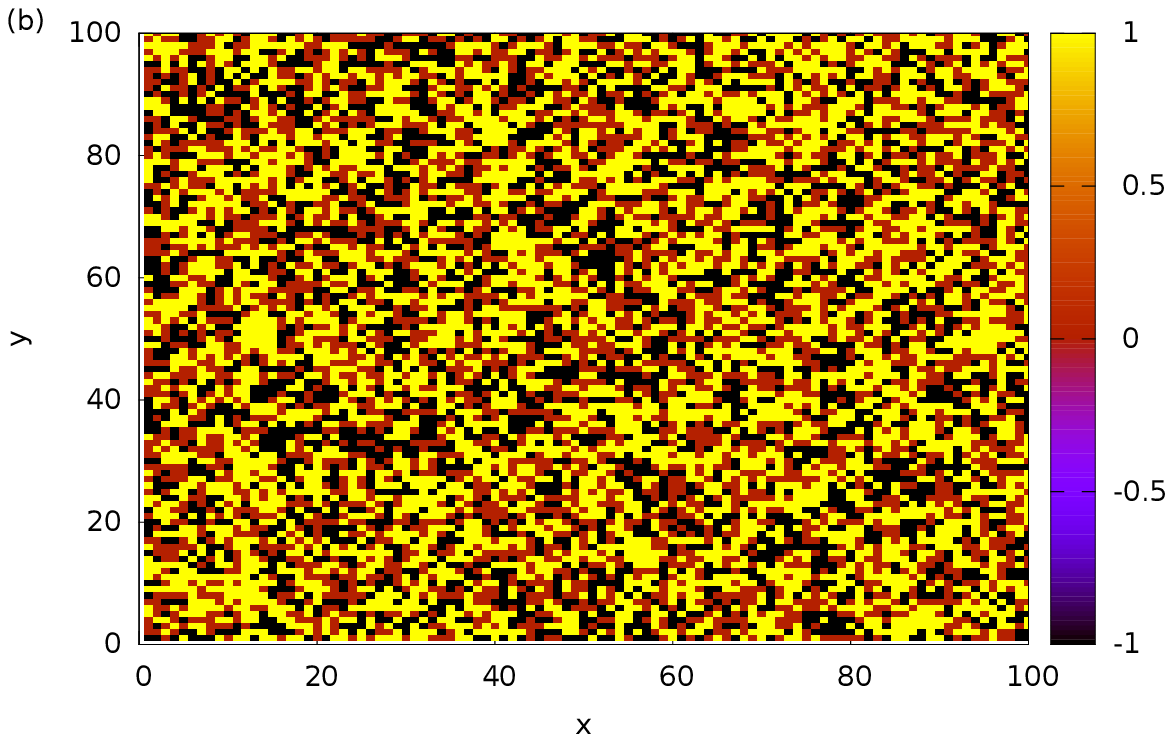}}
\\
\resizebox{7cm}{6cm}{\includegraphics[angle=0]{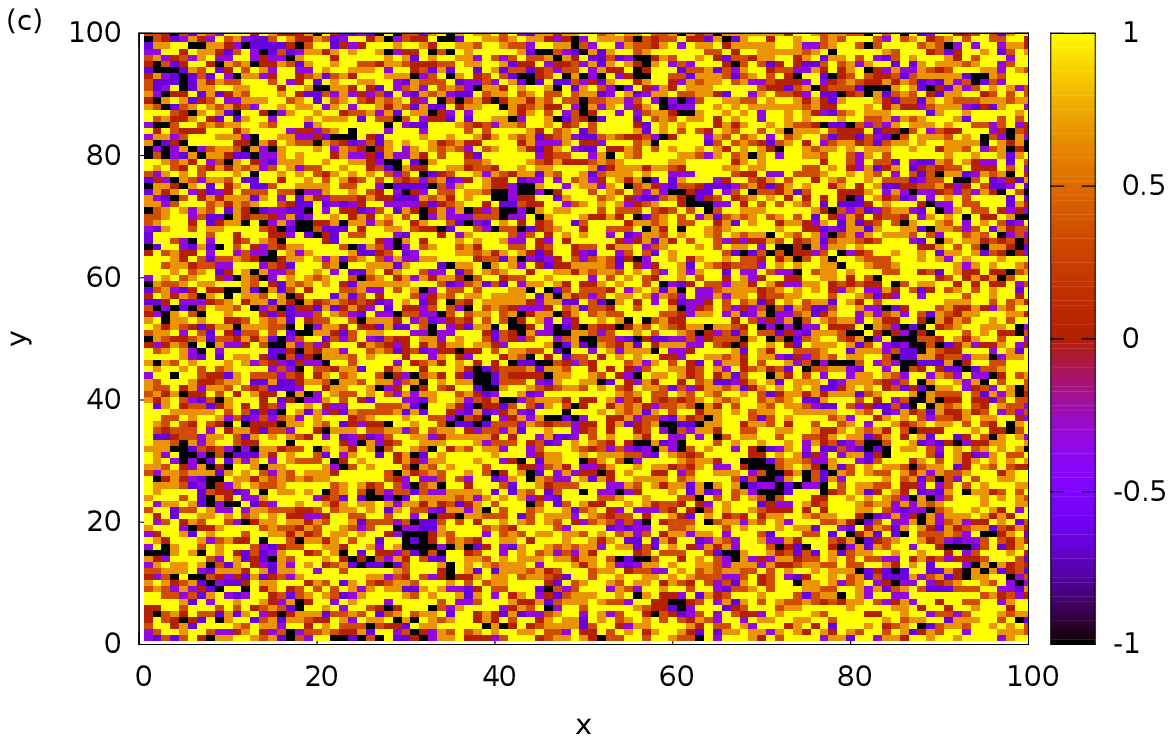}}
\resizebox{7cm}{6cm}{\includegraphics[angle=0]{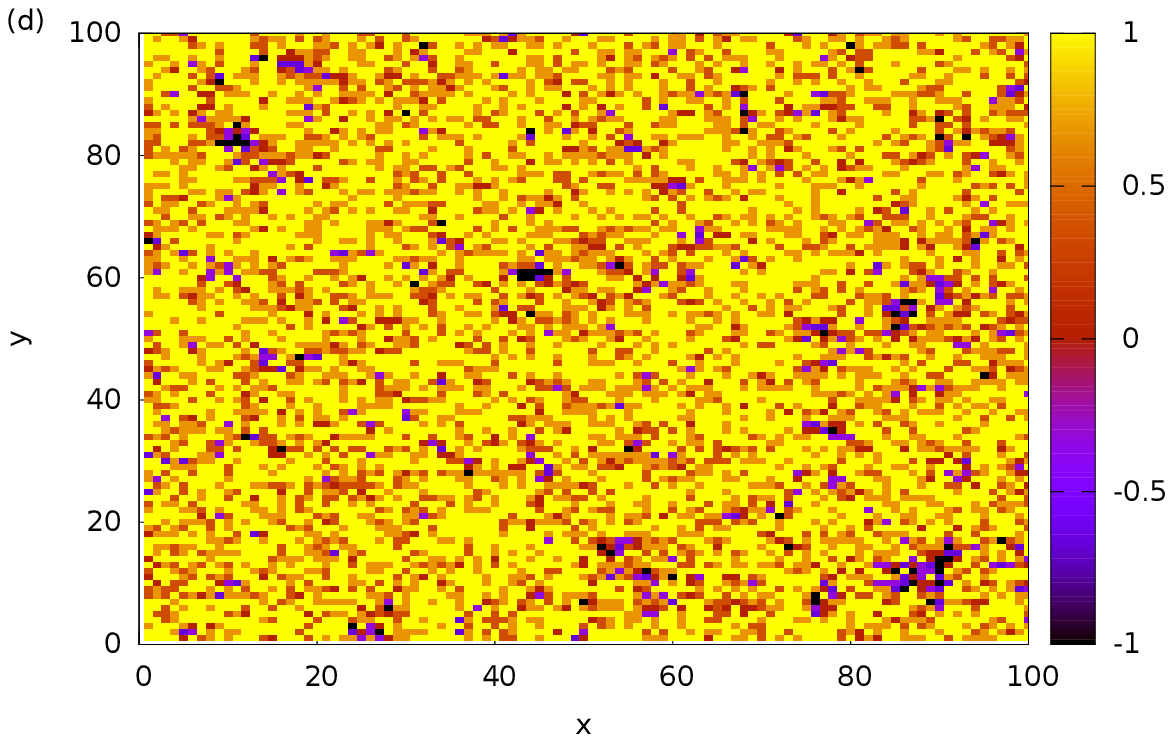}}
          \end{tabular}
\caption{\footnotesize{Lattice morphology (value of $S^z(x,y,t)$) 
for different values of $n$, $h_0$ and $T$ 
for uniformly oscillating magnetic field: 
(a) $n=5$, $h_0=0.8$ \& $T=0.8$ at time $t=3950$ and (b) $n=3$, $h_0=0.2$ \& $T=2.9$, 
(c) $n=7$, $h_0=0.3$ \& $T=1.5$, \& (d) $n=7$, $h_0=1.2$ \& $T=1.0$ at time $t=4000$.}}

\end{center}
\end{figure}

%%%%%%%%%%%%%%%%%%%%%%%%%%%%%%%%%%%%%%%%%%%%%%%%%%%%%%%%%%%%%%%%%%%%%%%%%%%%%%%%%%

\end{document}